\documentclass{article}

\usepackage{arxiv}
\usepackage{graphicx}
\usepackage[abs,unit=1pt]{overpic} 
\usepackage{newtxtext}
\usepackage{newtxmath}
\usepackage{natbib}
\usepackage{hyperref}
\usepackage{soul}
\hypersetup{
    colorlinks = true,
    urlcolor   = blue,
    citecolor  = black,
}

\newcommand{\RomanNumeralCaps}[1]
\linenumbers

\title{Separated flows over swept wings across\\ transitional Reynolds numbers}

\author{Laura Victoria Rolandi, Jonathan Quang Tran, and Kunihiko Taira\\
  Department of Mechanical and Aerospace Engineering\\
  University of California, Los Angeles, California 90095, USA}

\begin{document}
\date{}
\maketitle

\begin{abstract}
We explore the effect of wing sweep on the interaction between leading-edge separation and tip vortex for a finite wing with sweep angles $\Lambda=0^\circ$ to $45^\circ$, semi-aspect-ratio $sAR=2$, angle of attack $14^\circ$, and chord-based Reynolds numbers $Re=600$, $1000$, $2500$, $5000$, and $10000$.  Through this parametric study, we seek to bridge the knowledge gap between existing studies on separated flows over swept finite wings at low Reynolds numbers ($Re\approx 10^2$) and the behavior of turbulent flows at higher Reynolds numbers ($Re\approx10^4$). 
The flow structure significantly varies over this range of parameters, altering the lift characteristics. Specifically, increasing $\Lambda$ at $Re\leq5000$ reduces the lift, while increasing $\Lambda$ at $Re>5000$ enhances the lift. 
The flow modifications that lead to this effect with increasing Reynolds number are analyzed in terms of the streamwise vorticity, revealing three key effects: (i) a change in the sign of streamwise vorticity with increasing sweep angle, due to the higher spanwise velocity component, (ii) the progressive weakening and eventual disappearance of the tip vortex, when increasing the wing sweep, is accompanied by the emergence of a dominant inboard vortical structure near the root, and (iii)
 at higher Reynolds numbers and sweep angles, this inboard vortical structure merges with the main wake, intensifying the shear-layer roll-up and giving rise to a leading-edge vortex with an extended reattached region over the wing.
These findings reveal how the flow characteristics vary with Reynolds number, transitioning from regimes where swept wings offer limited benefit to regimes where they significantly enhance the aerodynamic performance. This study contributes to a deeper understanding of swept-wing wake dynamics, which is crucial for modern air vehicle design.
\end{abstract}

\section{Introduction}

The flow past low-aspect-ratio wings at high angle of attack exhibits complex three-dimensional, vortex-dominated wakes. Wingtip vortices generate significant downwash across a large portion of the span, producing nonlinear interactions between the tip vortices and the shed vortices. Introducing wing sweep further increases the complexity of the flow field. For backward-swept wings, flow unsteadiness tends to convect toward the wingtip, while additional vortical structures may emerge in the inboard region. The resulting interaction of these various mechanisms results in highly complex wake dynamics, particularly at high angles of attack.

The dynamics of unswept low-aspect-ratio wings have been studied both numerically \citep{taira2009three,zhang2020laminar,pandi2023streamwise,smith2024effect,rolandi2025triglobal}, and experimentally \citep{huang1995vortex,hayostek2018three,dong2020interplay,traub2021sweep} with focus on the effects of Reynolds number, wing geometry, and angle of attack \citep{Taira:ARFM27}. \citet{taira2009three} identified distinct flow regimes in finite wings at $Re = 300$ and $500$, highlighting transitions between stationary, periodic, and aperiodic wake structures over a range of angles of attack and wing aspect ratios. \citet{pandi2023streamwise} further explored the effects of aspect ratio and angle of attack at a Reynolds number of $Re=1000$. They found that interactions between the wing tip vortices and vortex shedding resulted in cellular wake shedding demarcated by various dislocation patterns, which was quantified by the variation of the spanwise vortex shedding frequency. Moreover, \cite{smith2024effect} investigated the interaction between the central wake and the tip vortex over a range of transitional Reynolds numbers. Their results showed that increasing the Reynolds number promoted instabilities in the leading-edge shear layer while also strengthening the wingtip vortex. Increasing Reynolds number is also associated with a widening of the shedding region in the wake, with small-scale structures spreading from the inboard toward the outboard regions of the wing.

While these studies provide insight into the wake dynamics of straight low-aspect-ratio wings, additional complexities are observed when wing sweep is introduced. Wing sweep promotes increased spanwise velocity over the wing, altering flow characteristics such as the boundary-layer separation and vortex formation \citep{harper1964review}. Studies on swept-wing aerodynamics indicate that increasing sweep generally has a stabilizing effect on the flow while significantly altering the wake topology partially owing to an interplay between reduced tip-vortex strength and increased spanwise flow intensity \citep{zhang2020laminar,ribeiro2023laminar,burtsev2024sensitivity}. As the sweep angle varies, interactions between the spanwise flow and the leading-edge vortex sheet can cause the separated shear layer to roll up over the wing surface, shifting the separation location and promoting the formation of prominent streamwise vortex structures that may ultimately lead to tip stall~\citep{black1956flow}. The various vortical structures that arise from wing sweep have been investigated at high Reynolds numbers \citep{breitsamter2008unsteady,schutte2017numerical} ($Re\approx 10^6$) as well as at moderate Reynolds numbers \citep{han2021leading} ($2.5\times 10^4 \lesssim Re \lesssim 52\times 10^5$).

For highly swept configurations, the flow can be accompanied by vortex lift, whereby vortices form and remain attached over the upper surface of the wing \citep{polhamus1966concept}. The low-pressure core of the attached vortex enhances suction, thereby increasing lift. At $Re=2.5\times10^4$, \cite{han2021leading} conducted experiments on a flat plate to investigate sweep effects ($0^\circ \le \Lambda \le 54^\circ$) for different wing aspect ratios ($2 \le AR \le 5$), and quantified the effect of sweep on leading-edge vortex lift. Their results showed that the theoretical decrease in lift from increased wing sweep was offset by an enhanced contribution from the leading-edge vortex. This was reflected by a gradual increase in the maximum lift coefficients as the sweep and aspect ratio were increased. Additionally, the contribution of the tip vortex to vortex lift has also been studied. For some wing configurations, the low pressure core of the tip vortex may extend over the wing suction surface, by which lift is increased \citep{lee2012vorticity,okamoto2019disappearance,hartlin2022sideslip,odaka2026extreme}.

\begin{figure} 
\centering
\includegraphics[width=0.65\textwidth]{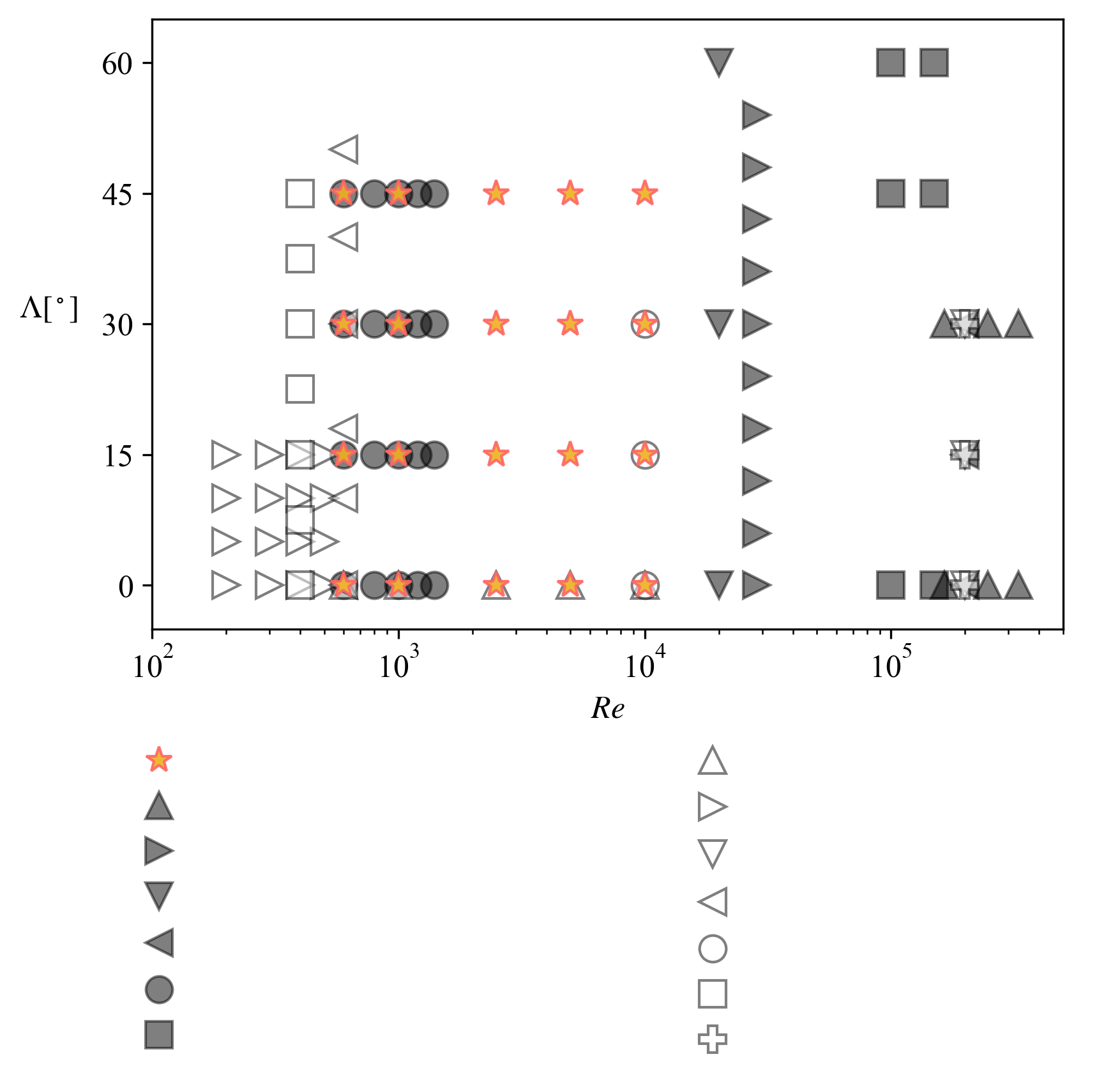}
\put(-251,87){Present study}
\put(-251,74.5){\cite{neal2026three}}
\put(-251,62){\cite{han2021leading}}
\put(-251,49.5){\cite{traub2021sweep}}
\put(-251,37){\cite{ullah2021experimental}}
\put(-251,24.5){\cite{hayostek2018three}}
\put(-251,12){\cite{traub2016experimental}}
\put(-100,87){\cite{smith2024effect}}
\put(-100,73.5){\cite{burtsev2024sensitivity}}
\put(-100,61){\cite{hammer2023effect}}
\put(-100,48){\cite{ribeiro2023laminar}}
\put(-100,35.5){\cite{zangeneh2021investigating}}
\put(-100,23){\cite{zhang2020laminar}}
\put(-100,11){\cite{visbal2019effect}}
\caption{Previous experimental (filled markers) and numerical (unfilled markers) studies of flow around a finite wing in the $(Re,\Lambda)$-plane. Note that the studies were performed across various aspect ratios, angles of attack, and pitching motions. \label{fig:Intro}}
\end{figure}
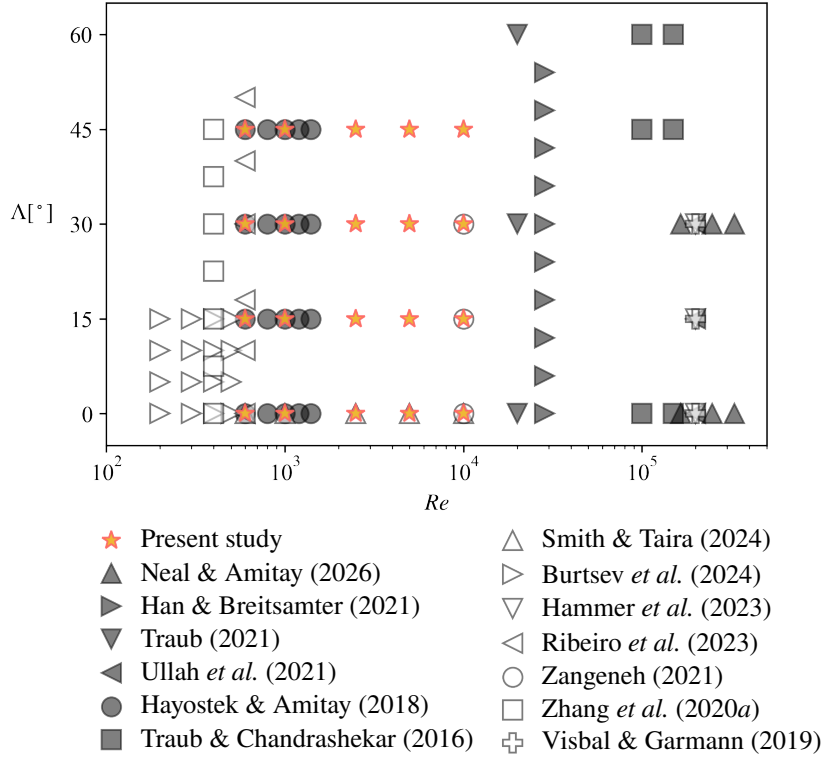

In Figure~\ref{fig:Intro}, we present the Reynolds numbers and sweep angle conditions investigated in the low to moderately high Reynolds number regimes. These previous studies considered different aspect ratios, angles of attack, and pitching motion and there is a critical gap in systematic studies of the effects of wing sweep on low-aspect-ratio wings over intermediate Reynolds numbers, specifically in the range between $Re\approx10^3$ to $10^4$. 
Over this range, in fact, the organized structures that are found in the wake at low Reynolds numbers start to lose their coherence \citep{huang1995vortex}. Analyzing this range is critical for understanding the mechanisms that bring about the formation of the attached leading-edge vortex and the beneficial effects of the vortex lift that is observed at high Reynolds numbers. In this work, we investigate the change in the flow dynamics over transitional Reynolds numbers, considering sweep angles up to $45^\circ$. We build upon the recent work by \cite{smith2024effect}, which considered flows around a low-aspect-ratio, unswept wing over the same Reynolds number range. 

The paper is organized as follows. In Sect. \ref{sec:Setup}, we introduce the computational setup and mesh convergence. The results are presented in Sect. \ref{sec:Results}, which consists of four parts. In Sect. \ref{sec:InstantaneousFlow} and Sect. \ref{sec:TiveAvg}, we analyze the unsteady flows and mean flows, respectively. An analysis of the ram's horn vortex is presented in Sect. \ref{sec:rams}, and in Sect. \ref{sec:AeroCoeff} we discuss the aerodynamic forces. Conclusions are offered in Sect. \ref{sec:Conclusions}.

\begin{figure} 
\centering
\includegraphics[width=\textwidth]{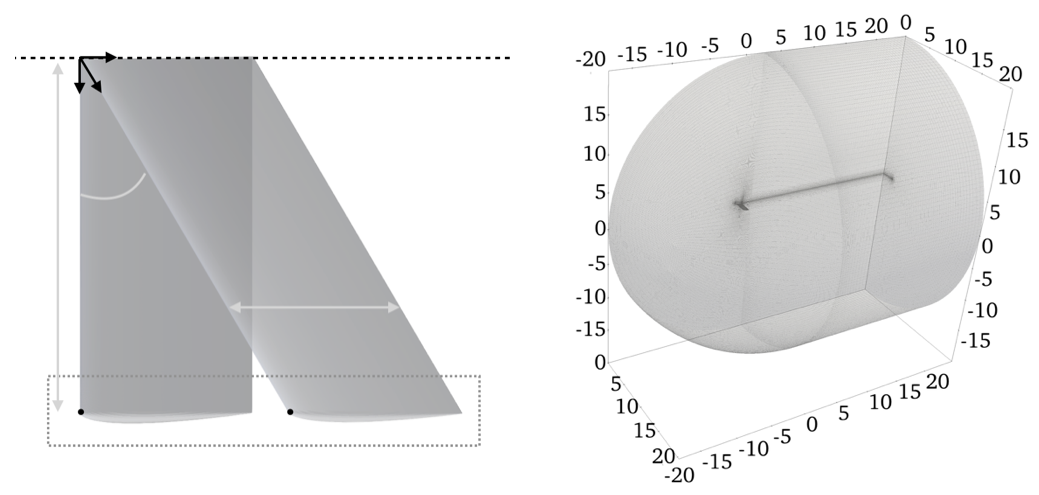}
\put(-450,210){(a)}
\put(-215,210){(b)}
 \put(-380,202){\scriptsize Symmetry plane}
     \put(-438,201){\scriptsize $O$}
     \put(-420,201){\scriptsize $x$}
  \put(-439,180){\scriptsize $z$}
  \put(-423,180){\scriptsize $z_\Lambda$}
\put(-421,138){\scriptsize $\Lambda$}
\put(-337,88){\scriptsize chord $c$}
\put(-380,25){\scriptsize rounded wingtip}
\put(-451,90){\rotatebox{90}{\scriptsize semi-span $b$}}
\put(-431,41){\scriptsize $(x_\text{tip},0,b)$}
\put(-338,41){\scriptsize  $(x_\text{tip},0,b)$}
\put(-100,20){\scriptsize $x$}
\put(-212,120){\scriptsize $y$}
\put(-200,30){\scriptsize $z$}
\caption{(a) Top view of the wing configuration and (b) computational domain.}\label{fig:Geometry}
\end{figure}
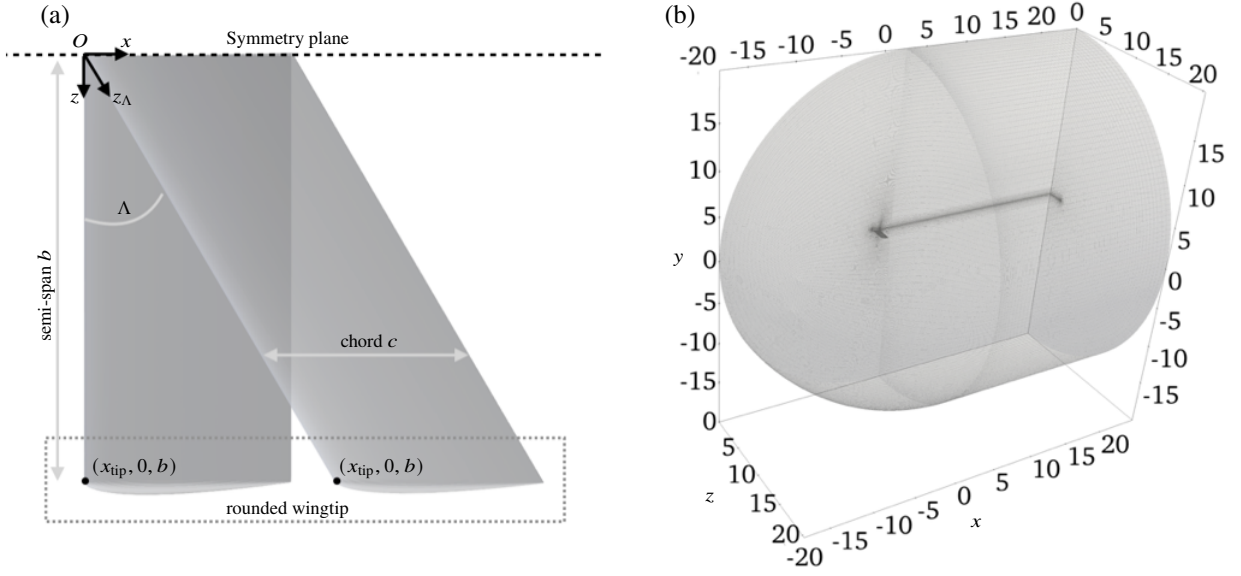

\section{Computational setup}\label{sec:Setup}
We study the flow over a NACA0012 cross-section finite wing with semi-aspect-ratio $sAR=b/c=2$, as shown in Figure~\ref{fig:Geometry}(a). Swept untapered wings are considered, with sweep angles of $\Lambda=0^\circ$, $15^\circ$, $30^\circ$, and $45^\circ$.  The sweep is applied after imposing the angle of attack on the finite wing, so that the leading edge at the root and at the tip lie in the same $xz$-plane. The origin is placed at the leading edge of the root. The streamwise location of the leading edge at the tip is indicated as $x_{\text{tip}}$, which varies with the sweep. 

We consider Reynolds numbers $Re\equiv U_\infty c/\nu=600,1000,2500,5000$ and $10000$, where $U_\infty$ is the freestream velocity magnitude and $\nu$ the kinematic viscosity, and Mach number $Ma=U_\infty/a=0.1$, where $a$ is the speed of sound. Direct numerical simulations (for $Re=600$, $1000$ and $2500$) and large eddy simulation (for $Re=5000$ and $10000$), using the Vreman subgrid scale model \citep{vreman2004eddy}, are performed with the compressible flow solver \textit{CharLES} \citep{bres2017unstructured,khalighi2011unstructured}. \textit{CharLES} uses a finite volume formulation with second-order accuracy in space and third-order accuracy in time, and the time step is selected such that the local acoustic Courant-Friedrichs-Lewy number is less than 1.

The non-dimensionalized spatial coordinates, which we henceforth denote just as $(x,y,z)$ for brevity, are normalized by the chord length $c$. The computational domain used in this work lies within $(x,y,z) \in [-20, 25] \times [-20,20] \times [0,20]$, as shown in Figure~\ref{fig:Geometry}(b), consistent with the setup in \citet{smith2024effect}. Details of the computational setup are provided in Table \ref{tab:gridRes}. In the table, $n_{\text{airfoil}}$ denotes the number of grid points along the airfoil profile, and $n_{z_{\Lambda}}$ represents the number of points distributed along the direction $\hat{e}_{z_\Lambda}=\hat{e}_x \sin \Lambda+ \hat{e}_z \cos \Lambda $, spanning a total length of $sAR / \cos \Lambda$. 

For the time-averaged flow fields and quantities shown in this paper, we compute the temporal average after the initial transient for a duration of $T_{\text{avg}} \ge 80 U_\infty t/c$ for $Re\leq1000$, and $T_{\text{avg}} \ge  150 U_\infty t/c$ for $Re\geq2500$. 
Over the last $20 U_\infty t/c$ of the averaging interval, the cumulative mean of the lift coefficient was seen to vary less than 2\% of its final value.
Throughout this work, time-averaged quantities are denoted by $\overline{(\,\cdot\,)}$.

\begin{table}
  \begin{center}
\def~{\hphantom{0}}
  \begin{tabular}{cccccccccc}
                   $Re$    &  $n_{\text{airfoil}}$ &  $n_{z_{\Lambda_{0}}}$ &  $n_{z_{\Lambda_{15}}}$ & $n_{z_{\Lambda_{30}}}$ & $n_{z_{\Lambda_{45}}}$ &  Cell count$_{\Lambda_{0}}$ &  Cell count$_{\Lambda_{15}}$&  Cell count$_{\Lambda_{30}}$&  Cell count$_{\Lambda_{45}}$ \\[3pt]
               600   &  100 & 50  &65 & 70& 85 & $4.0\times10^6$ & $7.9\times10^6$&  $9.1\times10^6$&  $12.4\times10^6$ \\[3pt]
              1000  &  120 & 75 & 80 & 85 &105 & $11.4\times10^6$ & $15.7\times10^6$& $17.9\times10^6$& $24.9\times10^6$\\[3pt]
                2500  &  120&  75 & 80 & 85 & 105 & $11.4\times10^6$ & $15.7\times10^6$ &$17.9\times10^6$&$24.9\times10^6$\\[3pt]
                 5000  &  160 & 105 & 110 &120 &150 &$26.5\times10^6$ & $33.5\times10^6$& $39.6\times10^6$& $59.9\times10^6$\\[3pt]
                  10000  & 160  &  105 & 110 &120 &150 &$26.5\times10^6$ & $33.5\times10^6$& $39.6\times10^6$& $59.9\times10^6$\\
  \end{tabular}
  \caption{Mesh details for the different Reynolds numbers and sweep angles.}
  \label{tab:gridRes}
  \end{center}
\end{table}

\section{Results}\label{sec:Results}

\begin{figure} 
\centering
\includegraphics[width=\textwidth]{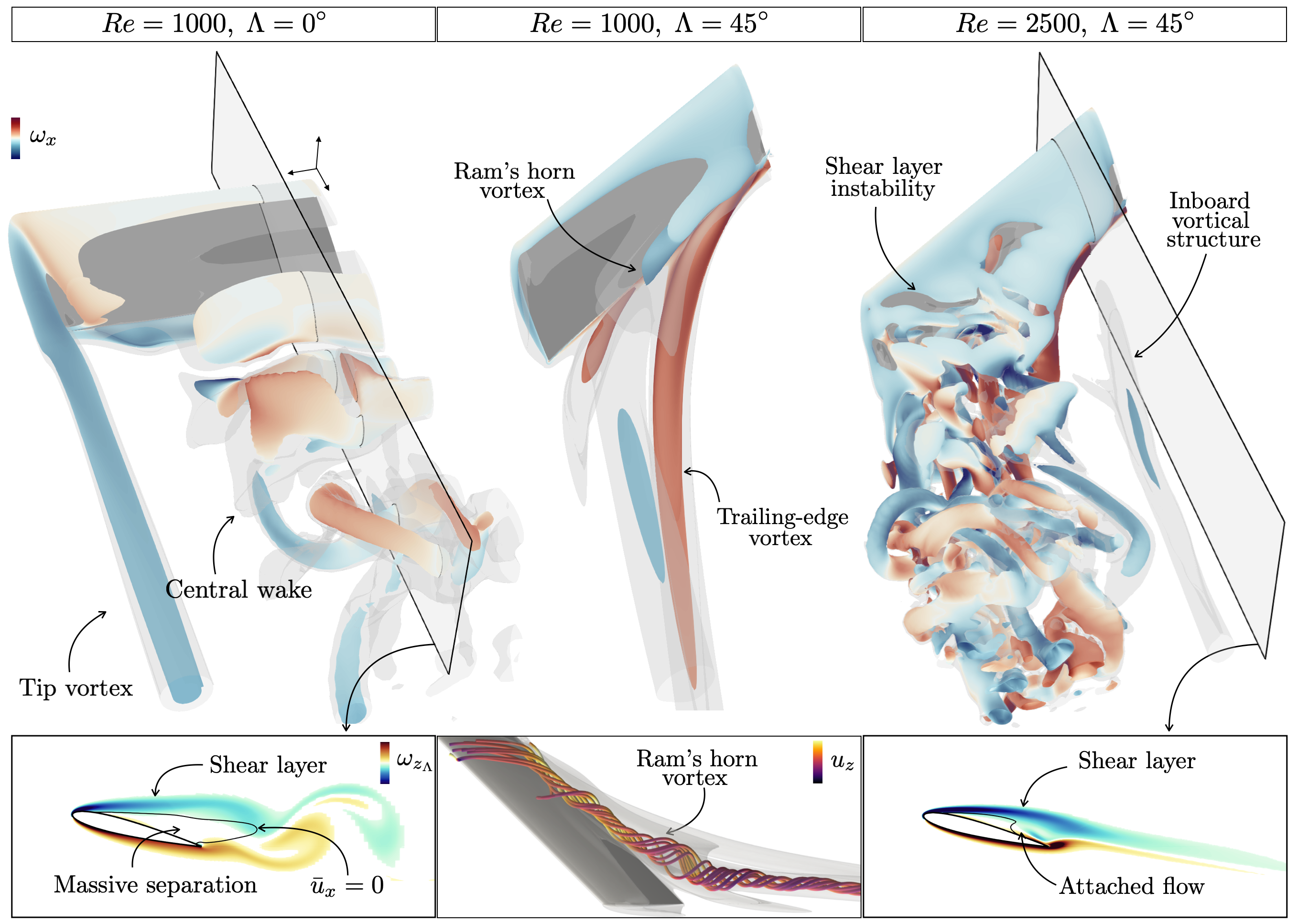}
\put(-349,268){\tiny{$x$}}
\put(-352,281){\tiny{$y$}}
\put(-365,276){\tiny{$z$}}
\caption{Illustration of flow structures for three representative cases visualized using Q-criterion isosurface. Bottom insets present complementary spanwise cross-sectional and localized visualizations of the flow. \label{fig:IntroFlowField}}
\end{figure}

Let us first introduce and describe the key flow features relevant to our discussion.  Flow structures from three representative cases are shown in Figure~\ref{fig:IntroFlowField}. At low Reynolds numbers and low sweep angles, illustrated here by the case with $Re=1000$ and $\Lambda=0^\circ$, the flow is characterized by two main features: a coherent tip vortex, which forms a steady elongated structure near the wingtip, and a central unsteady wake originating near the wing root. The tip vortex forms when flow from the pressure side of the wing rolls around the wingtip towards the suction side, generating a streamwise-oriented vortex that persists downstream in the wake.
The central wake is developed near the root and consists of counter-rotating spanwise vortices that are shed downstream, as shown in the bottom inset of Figure~\ref{fig:IntroFlowField}. At sufficiently high angles of attack, the boundary layer separates from the airfoil surface near the leading edge forming a shear layer above the wing. As the shear layer convects downstream, it rolls up leading to periodic vortex shedding. 
Due to the high viscosity and the consequent thickness of the boundary layer at low Reynolds numbers, the shear layer rolls up far downstream from the separation point. This results in a massive separation region on the suction side of the airfoil, as indicated by the $\bar{u}_x=0$ contour.

At high sweep angles, as illustrated by the case at $Re=1000$ and $\Lambda=45^\circ$ in Figure~\ref{fig:IntroFlowField}, the flow develops a vortical structure that originates near the wing root and extends into the wake behind the tip. This vortical structure is sometimes referred to as the \textit{ram's horn vortex} \citep{black1956flow} or as a spiral vortex \citep{poll1986spiral}. 
The ram's horn vortex is generated through the roll-up of the leading-edge shear layer near the root over the suction side.
The vortex emerges together with a strong counter-rotating vortex emanating from the trailing edge and is initially oriented along the $z_\Lambda$ direction near the root. As it convects downstream, the freestream progressively tilts it into a streamwise-oriented vortex that persists in the wake. The streamlines in the bottom inset indicate that this vortical structure is fed by flow coming from the leading edge at the root and show the reorientation at the wingtip.

At higher Reynolds numbers and high sweep angles, as illustrated by the case at $Re=2500$ and $\Lambda=45^\circ$ in Figure~\ref{fig:IntroFlowField}, the unsteady wake is shifted towards the outboard part of the wing. The increase in Reynolds number leads to a thinner boundary layer that detaches from the wing forming a thin shear layer. The thin shear layer develops short-wavelength instabilities in the outboard region, feeding the streamwise-oriented wake. Moreover, the thin shear layer rolls up at the root closer to the wing compared to unswept cases at lower Reynolds numbers, promoting the emergence of a region of attached flow on the wing, as indicated by the $\bar{u}_x=0$ contour. This case also reveals the emergence of an inboard vortical structure. This inboard vortical structure is quasi-steady and is characterized by negative streamwise vorticity, similar to that generated by the shear layer, indicating that it originates from the flow above the wing.

Now that we have described the key flow features relevant to our discussions, let us further describe our results. We begin by examining the instantaneous characteristics of the flow, identifying unsteadiness, and then proceed to analyze the time-averaged features, with a focus on the development of vortical structures over the Reynolds numbers and sweep angles.

\subsection{Unsteady dynamics}\label{sec:InstantaneousFlow}

\begin{figure} 
\centering
\includegraphics[width=\textwidth]{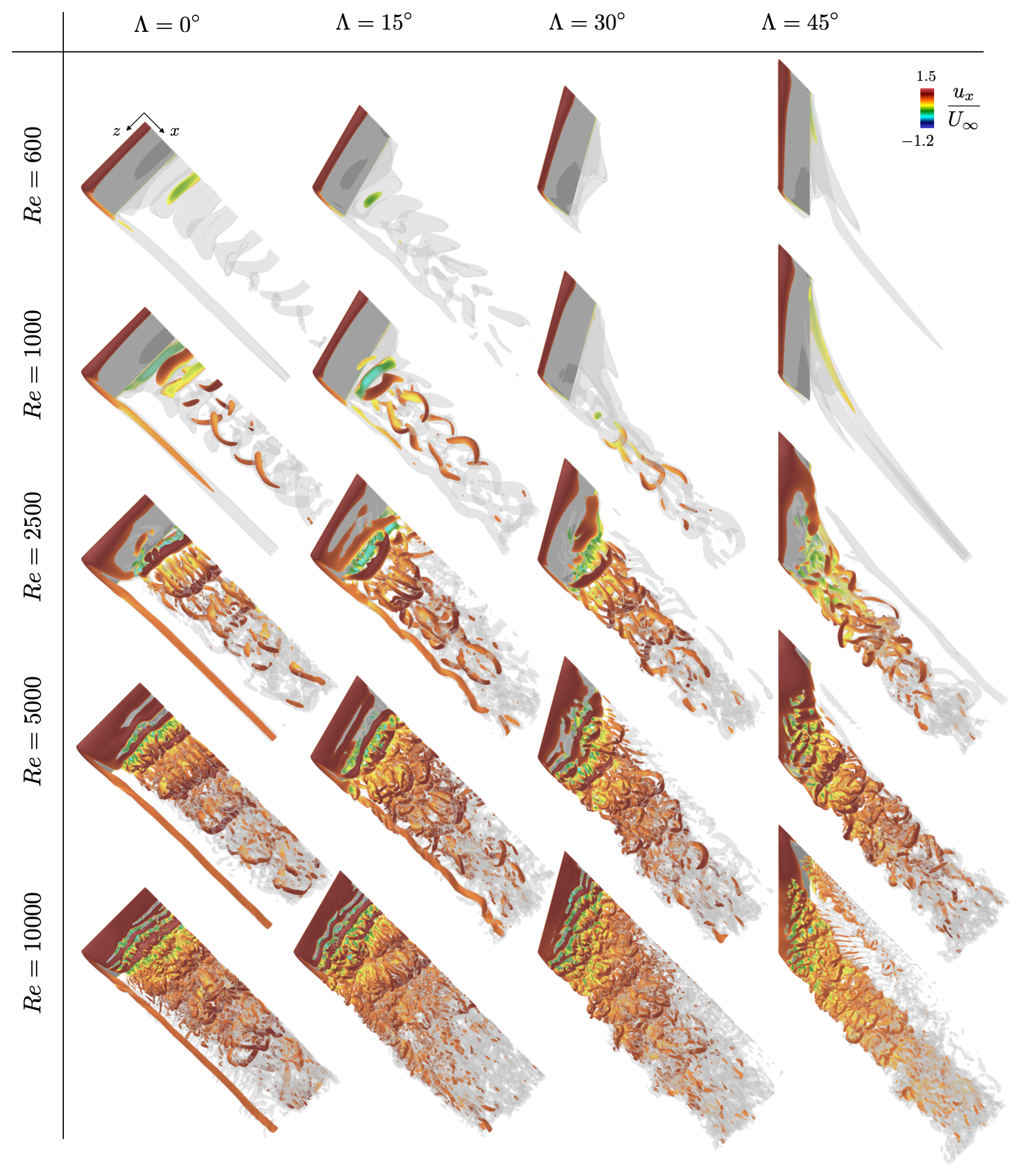}
\caption{Instantaneous flow fields around a NACA0012 finite wing at different Reynolds numbers and sweep angles. Visualization with the isosurface of Q-criterion $Qc^2/U_{\infty}^2=5$ colored by streamwise velocity superposed on translucent isosurface of Q-criterion $Qc^2/U_{\infty}^2=0.5$. \label{fig:FlowField}}
\end{figure}

Let us first identify dominant trends in the unsteady dynamics. We show representative instantaneous flow fields for the different sweep angles and Reynolds numbers considered in Figure~\ref{fig:FlowField}. First, increasing sweep introduces spanwise velocity that shifts the dominant unsteadiness from the inboard region toward the tip, weakening the tip vortex and stabilizing the inboard flow. Sweep has an overall attenuating effect on flow fluctuations, fully stabilizing the flow at low Reynolds numbers and generally reducing fluctuation amplitudes at higher Reynolds numbers, consistent with previous results \citep{zhang2020laminar,ribeiro2023laminar,burtsev2024sensitivity}. Secondly, increasing the Reynolds number generally increases the spanwise extent of the unsteady wake region across all sweep angles due to decreased stability of the shear layer. For unswept wings, this expansion occurs from the inboard central wake towards the tip \citep{taira2009three,smith2024effect}, whereas for swept wings the unsteady region extends from the outboard portion of the wing towards the root. 
Overall, we see a complex interplay between the stabilizing effect of increasing sweep and the destabilizing effect of increasing Reynolds number leading to the variety of flow responses observed.

For the unswept cases ($\Lambda = 0^\circ$) the flow exhibits both a coherent tip vortex and a central unsteady wake originating near the wing root. At all Reynolds numbers, these two structures remain clearly segregated, as the downwash induced by the tip vortex stabilizes the leading-edge shear layer \citep{taira2009three}. As the Reynolds number increases, however, fluctuations from the central unsteady wake extend progressively towards the tip, reducing the separation between the two regions \citep{smith2024effect}. At higher Reynolds numbers, the thinner shear layer becomes more receptive to instabilities, leading to roll-up of the vortex sheet above the wing \citep{huang1995vortex}, even in the presence of the tip vortex. 

With the introduction of sweep, the unsteady wake structures associated with the central wake for the unswept case are generally displaced away from the root region toward the outboard. This is a direct consequence of the spanwise velocity induced by wing sweep, which transports structures away from the root plane. As a result, the most prominent unsteady wake structures progressively concentrate near the tip as the sweep angle is increased. The only exception occurs at sufficiently low Reynolds numbers (e.g., $Re=600$), where the increased sweep fully stabilizes the flow. 
Additionally, the elongated spanwise structures associated with the roll-up of the shear layer appear to be oriented along the direction of sweep.

We observe that wing sweep promotes interactions between the flow around the tip and the unsteady shedding in the wake. For example, in the $Re=2500$ cases, at low sweep angles ($\Lambda=15^\circ$) the unsteady wake induces helical oscillations in the axis of the tip vortex as it persists downstream. Keeping the Reynolds number fixed and increasing sweep angle ($\Lambda\ge30^\circ$), the unsteady wake completely disrupts the formation of the tip vortex as unsteadiness in the flow is increasingly concentrated behind the outboard tip region.

On the other hand, when keeping the sweep angle fixed and increasing Reynolds number, we see that even low sweep angles are sufficient to completely disrupt the tip vortex formation. This is seen most clearly for the $\Lambda=15^\circ$ cases. For $(Re,\Lambda)=(1000, 15^\circ)$ two series of braid-like vortices are observed which produce slight oscillations in the tip vortex, similar to those observed at lower Reynolds number and larger AR \citep{burtsev2022linear}. Increasing the Reynolds number, the frequency and intensity of the wake structures increase, and at $Re=5000$ they start to twist and break apart the tip vortex. At the highest Reynolds number, $Re=10000$, the tip vortex is broken down completely into small scale structures. At the same time, the spanwise extent of the unsteady wake originating from the outboard region of the wing steadily expands towards the root with increasing Reynolds number. This occurs because the stabilizing effect that sweep has on the shear layer near the root is counteracted by the thinning of the shear layer at higher Reynolds numbers. Additionally, the thinner shear layer begins to roll up and break down closer to the leading edge as Reynolds number increases.

At the highest sweep angle, $\Lambda=45^\circ$, the flow remains stationary for the $Re=600$ and $1000$ cases. For this sweep angle, we can observe the ram's horn vortex in the unsteady flow fields for these stationary cases. At $(Re,\Lambda) = (2500, 45^\circ)$, the ram's horn vortex is no longer explicitly visible in the unsteady flow, but instead there is an emergence of an elongated, quasi-steady, streamwise aligned vortex that persists downstream behind the inboard region of the wing (the so-called ``inboard vortical structure'' in Figure~\ref{fig:IntroFlowField}). This structure is not clearly distinct at $Re=5000$. At $(Re,\Lambda) = (10000, 45^\circ)$, the leading-edge shear layer reattaches near the root, forming a leading-edge vortex \citep{eldredge2019leading}, which extends over the inboard region of the wing. This is similar to observations made of the flow over a swept flat plate at a similar $sAR$ \citep{han2021leading}.
Additionally, the $(Re,\Lambda) = (10000, 45^\circ)$ case sees the emergence of finer vortical structures in the wake close to the root, which are distinctly separated from the large unsteady wake that starts at the tip.

\begin{figure}
\centering
\begin{overpic}[width=\textwidth]{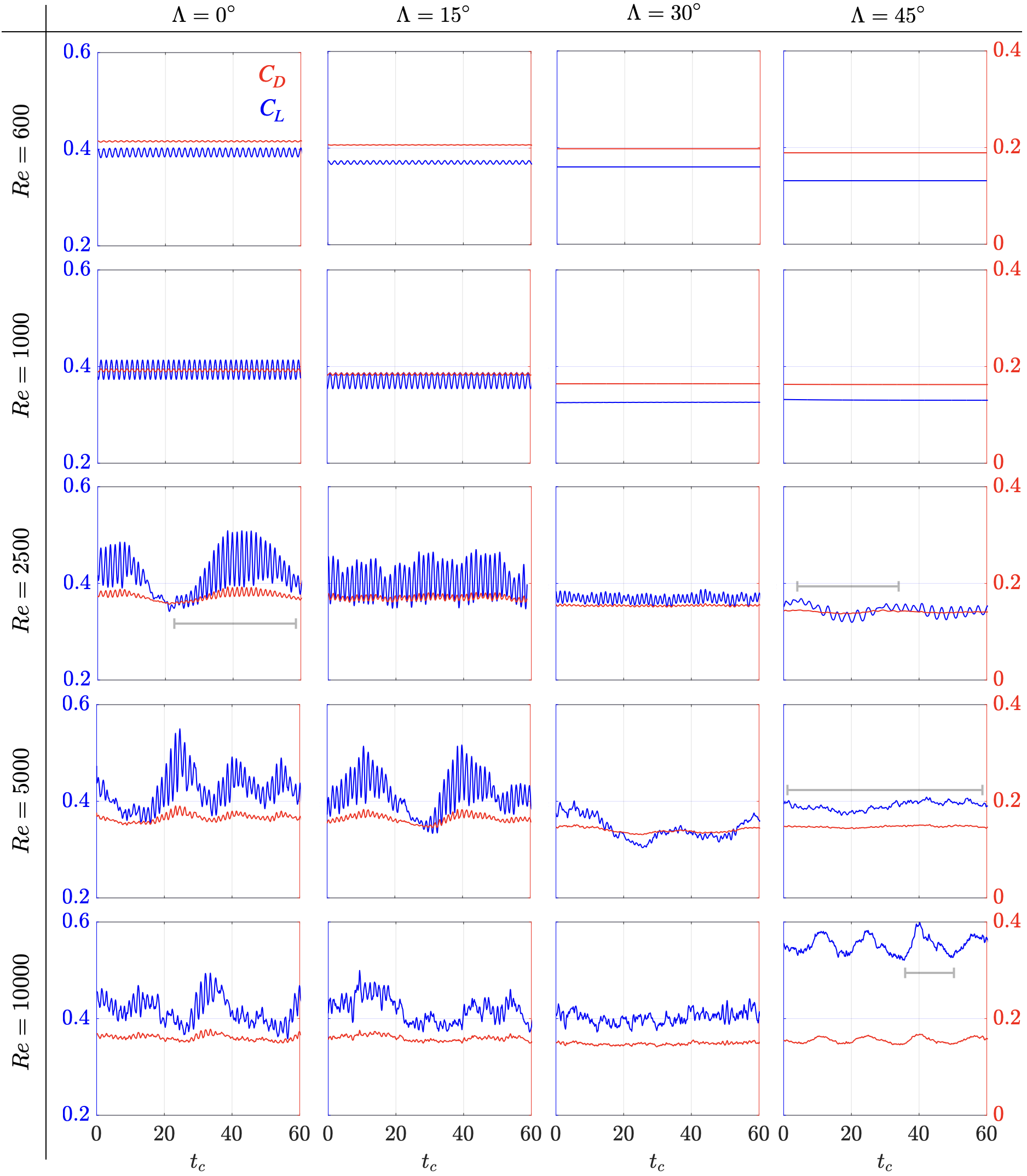}
  \put(100,245){\tiny{$\approx40 t_c$}}
  \put(380,275){\tiny{$\approx30 t_c$}}
  \put(400,180){\tiny{$\approx60 t_c$}}
  \put(418,85){\tiny{$\approx15 t_c$}}
  \put(260,500){\tiny{Stationary}}
  \put(370,500){\tiny{Stationary}}
  \put(370,400){\tiny{Stationary}}
\end{overpic}
\caption{Instantaneous lift ($C_L$) and drag ($C_D$) coefficients. \label{fig:Forces}}
\end{figure}
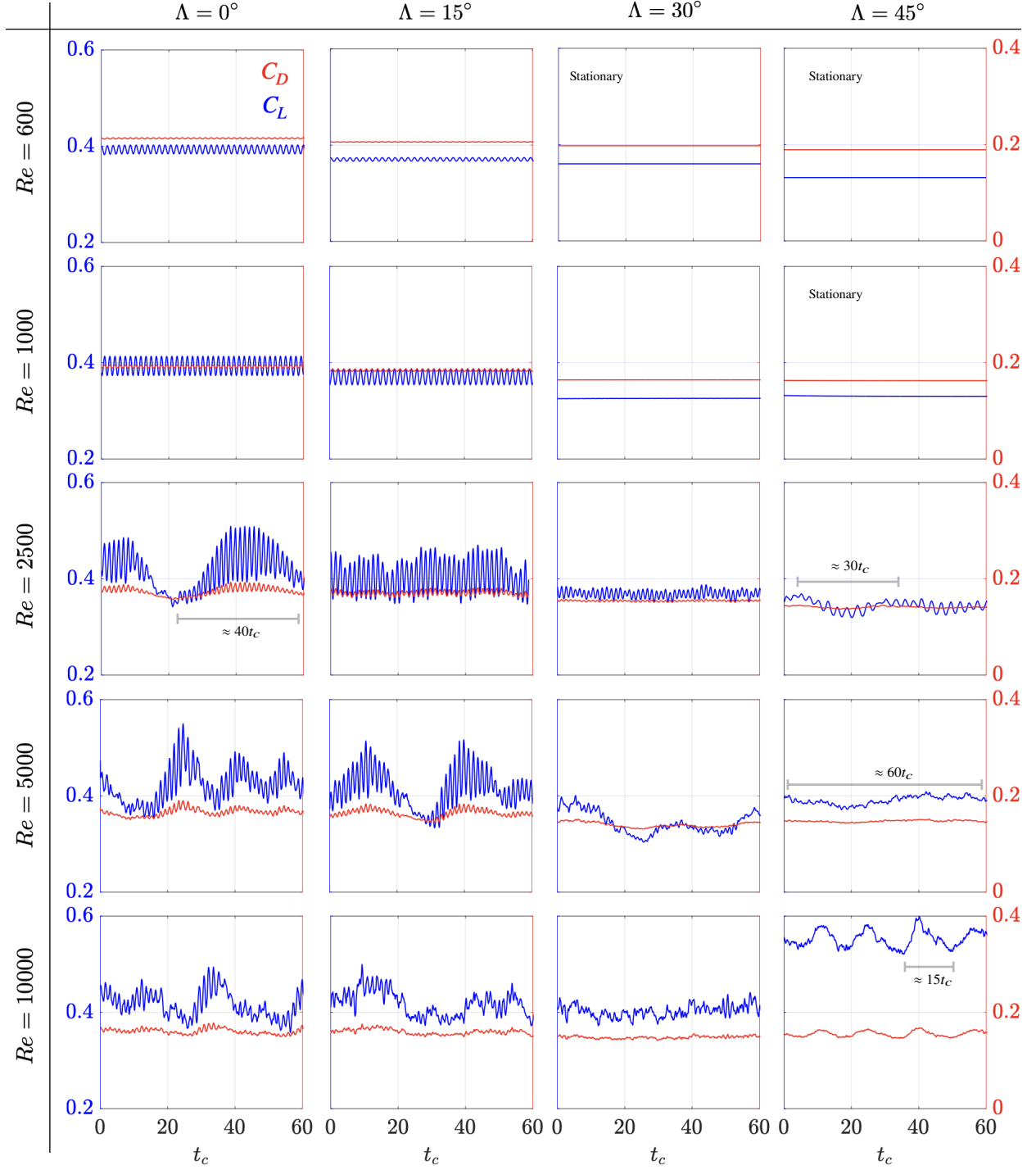

The instantaneous aerodynamic forces acting on the wing can provide further insight into the unsteady dynamics. The time history of lift and drag coefficients for the different cases are presented in Figure~\ref{fig:Forces}. The lift coefficient exhibits larger amplitude oscillations compared to the drag coefficient, however their temporal trends appear similar. While we discuss the unsteady instantaneous forces here, a detailed discussion of the mean values of the lift and drag coefficients is presented in Sect. \ref{sec:AeroCoeff}.

At $Re=600$ and $1000$, the forces are periodic in time. Increasing the sweep stabilizes the flow, ultimately resulting in stationary aerodynamic forces for these low Reynolds numbers. We note that while the forces for the $(Re,\Lambda)=(1000, 30^\circ)$ case appear stationary, there are oscillations in the flow field that occur downstream, which lead to negligibly small oscillations in the aerodynamic coefficients. 

Considering just the unswept cases, when Reynolds number is increased to $Re=2500$, the aerodynamic forces exhibit higher amplitude oscillations due to the increased vortex strength. Additionally, the forces exhibit a lower-frequency envelope, similar to that reported for spanwise-periodic wings \citep{rolandi2025biglobal}. This behavior is attributed to the ``breathing'' of the separation region, during which the leading-edge shear layer roll-up oscillates toward and away from the airfoil surface. For the higher Reynolds number cases ($Re = 5000$ and $10000$), there is increased intermittency in the aerodynamic forces, reflective of the loss of coherence of vortical structures \citep{huang1995vortex}. 

When the sweep angle is increased for $Re \ge 2500$, while the flow is not completely stabilized, we generally see attenuation in the oscillation amplitude of the aerodynamic forces. However, there are some notably different behaviors seen for the highest sweep angle. A low-frequency envelope reemerges in the forces, which corresponds to oscillations of structures in the inboard wake, such as the elongated, streamwise aligned, inboard vortical structure seen in the $(Re,\Lambda) = (2500, 45^\circ)$ case in Figure~\ref{fig:FlowField}. For the highest sweep angle and Reynolds number, $(Re,\Lambda) = (10000, 45^\circ)$, there is a substantial increase in the lift. This is generated by vortex lift associated with the rolling up of the leading edge shear layer closer to the suction side near the root, which we will further discuss in later sections.

\begin{figure} 
\centering
\includegraphics[width=0.78\textwidth]{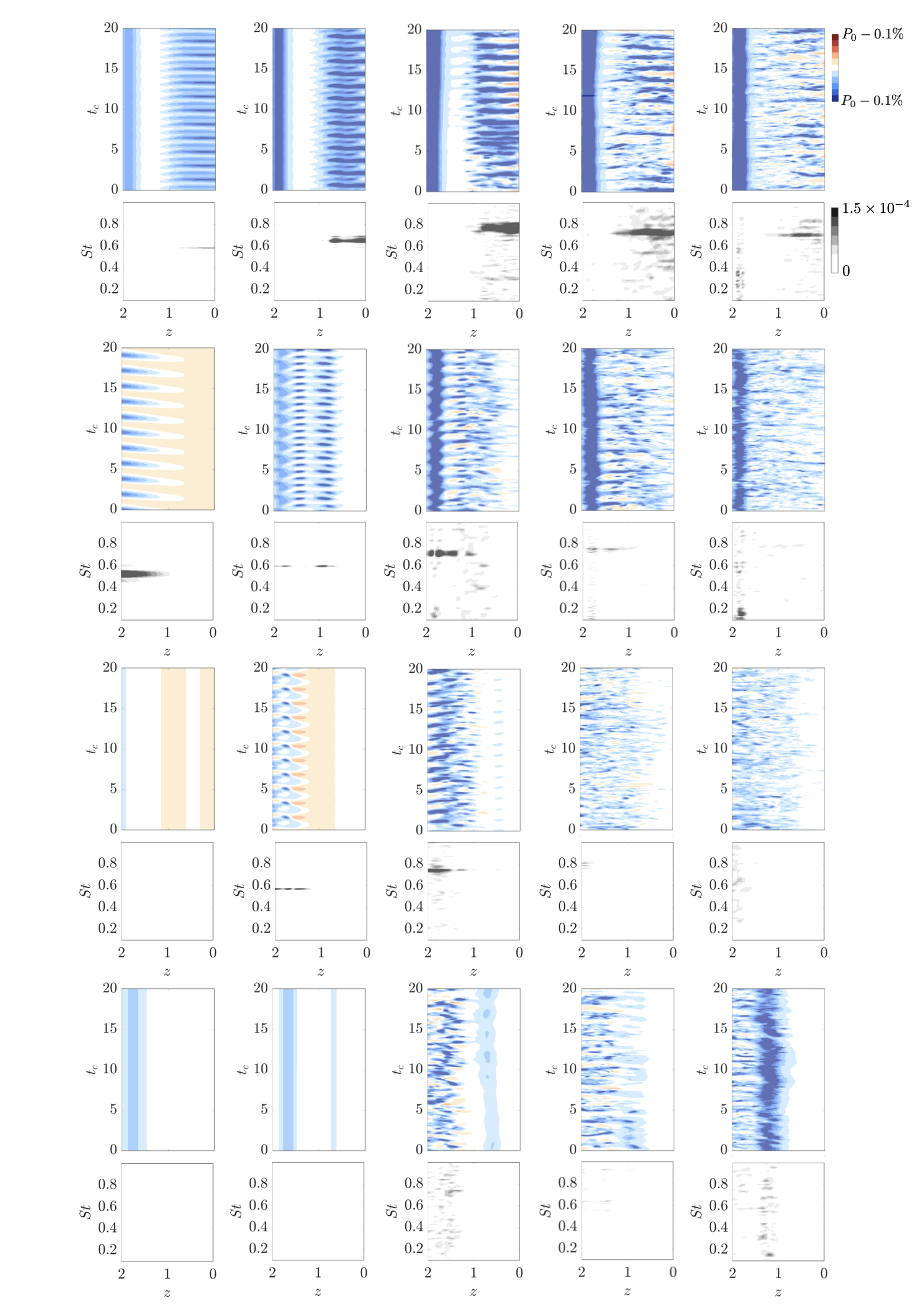}
\put(-318,515){\rotatebox{0}{$Re=600$}}
\put(-260,515){\rotatebox{0}{$Re=1000$}}
\put(-200,515){\rotatebox{0}{$Re=2500$}}
\put(-138,515){\rotatebox{0}{$Re=5000$}}
\put(-80,515){\rotatebox{0}{$Re=10000$}}
\put(-360,440){{$\Lambda=0^\circ$}}
\put(-360,315){{$\Lambda=15^\circ$}}
\put(-360,188){{$\Lambda=30^\circ$}}
\put(-360,60){{$\Lambda=45^\circ$}}
\caption{The instantaneous pressure (top) and the pressure spectra (bottom) at $(x,y)=(x_{\text{tip}}+3,-0.24)$ for the different Reynolds numbers and sweep angles. \label{fig:Press}}
\end{figure}

We next examine how changes in the flow dynamics are reflected in the wake pressure signal. Figure~\ref{fig:Press} presents the pressure measured along a line downstream of the trailing edge, extending from $(x_{\text{tip}} + 3, -0.24, 0)$ to $(x_{\text{tip}} + 3, -0.24, 2)$. The vertical position $y=-0.24$ indicates the $xz$-plane that contains the trailing edge. The figure also includes the corresponding pressure spectra evaluated across the span.

For the unswept case ($\Lambda = 0^\circ$), both the tip vortex and the unsteady central wake are clearly evident in the pressure signature across all Reynolds numbers. At low Reynolds numbers, the wake dynamics are largely periodic, resulting in a single dominant peak in the frequency spectrum. In particular, at $Re = 1000$, the peak at $St \approx 0.7$ closely matches the shedding frequency reported for a spanwise-periodic wing at the same Reynolds number \citep{rolandi2025biglobal}. As the Reynolds number increases, the pressure signal becomes less coherent, leading to a broader frequency spectrum. This behavior is consistent with the previously noted intermittency in the force response. Additionally, as the central wake shifts closer to the tip region, the tip vortex begins to exhibit low-frequency oscillations. 

As the sweep angle increases, the dominant peak in the pressure spectra is located toward the tip region across all Reynolds numbers. An exception occurs at low Reynolds numbers and the highest sweep angles, where the pressure signal becomes nearly steady due to sweep-induced flow stabilization.
When $Re \leq 2500$, increasing sweep initially results in a slight decrease of the dominant frequency (e.g., see the $Re = 1000$, $0^\circ \le \Lambda \le 30^\circ$ cases), while the dominant frequency remains relatively unchanged for the $Re = 5000$ cases. 
For the highest Reynolds number case ($Re = 10000$), increasing the sweep from $\Lambda = 0^\circ$ to $15^\circ$ leads to a significant reduction in the dominant frequency, from $St \approx 0.75$ in the central wake to $St \approx 0.2$ near the tip. In all cases where a tip vortex is present, the pressure within the vortex decreases with increasing Reynolds number, indicating a strengthening of the tip vortex.
Overall, increasing both sweep angle and Reynolds number leads to a loss of coherence in the wake, resulting in broader and reduced spectral content in the pressure signal.

\begin{figure} 
\centering
\includegraphics[width=\textwidth]{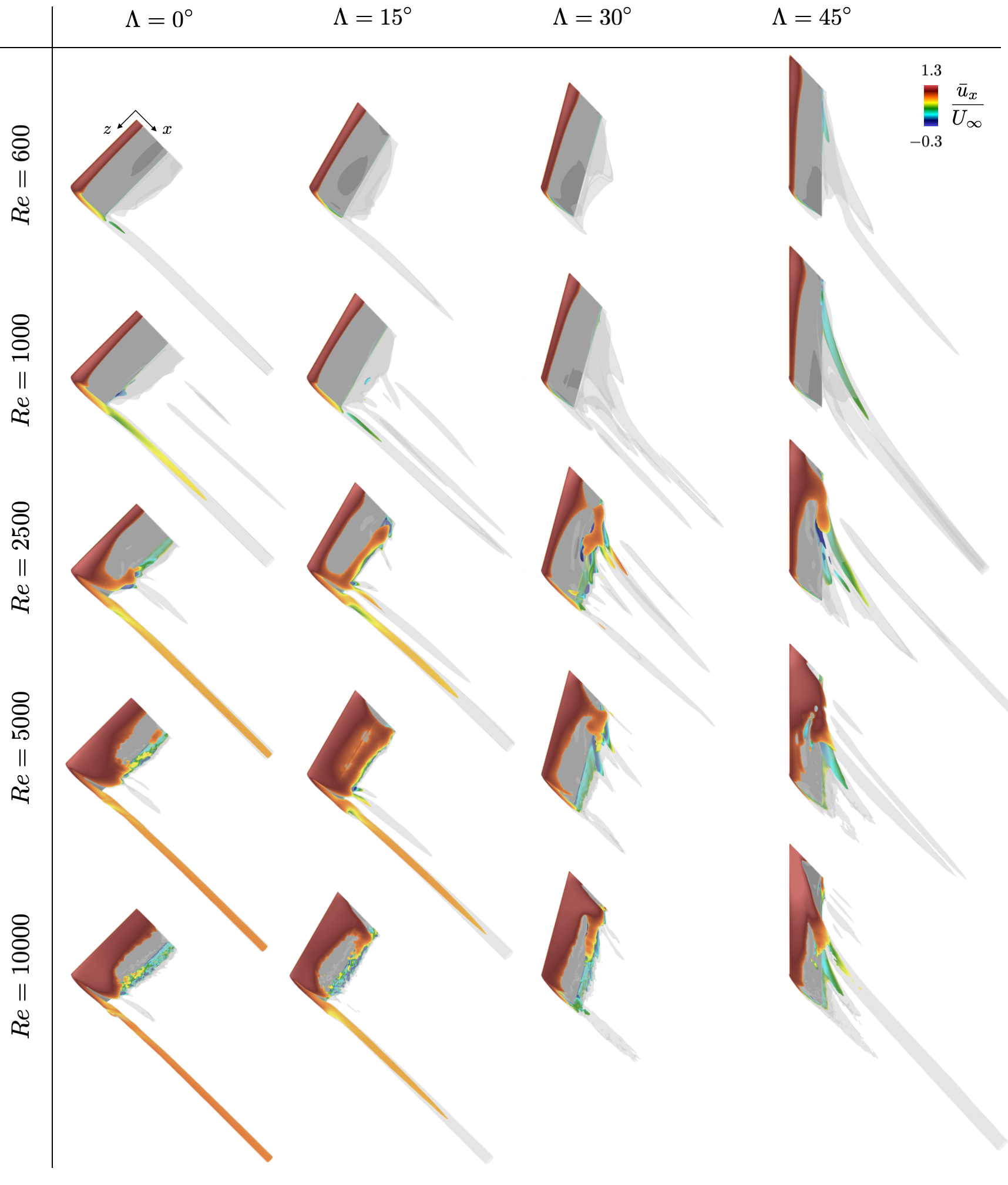}
\caption{Time-averaged flow fields around a NACA0012 finite wing at different Reynolds numbers and sweep angles. Visualization with the isosurface of Q-criterion $Qc^2/U_{\infty}^2=5$ colored by time-averaged streamwise velocity superposed on translucent isosurface of Q-criterion $Qc^2/U_{\infty}^2=0.5$. \label{fig:FlowFieldAVG}}
\end{figure}

\subsection{Mean flow vortical structures}\label{sec:TiveAvg}

The time-averaged flow field highlights regions of flow separation and structures that persist over time. We present the time-averaged flow fields at the different sweep angles and Reynolds numbers considered in Figure~\ref{fig:FlowFieldAVG}. As in the case with the instantaneous flow fields in Figure~\ref{fig:FlowField}, these fields vary significantly across the parameter space. 

While case-specific nuances exist, several overall trends are evident in the mean flows shown in Figure~\ref{fig:FlowFieldAVG}. As observed in the instantaneous flow fields, increasing the sweep angle promotes the breakdown of the tip vortex. In addition, sweep gives rise to elongated vortical structures that persist downstream (in the time-average sense). These structures originate from the roll-up of the leading-edge shear layer near the root and are eventually turned in the direction of the free stream, such as the aforementioned ram's horn vortex. 

One of the most prominent features in the time-averaged flow fields is the tip vortex. For both the unswept and $\Lambda = 15^\circ$ cases, the tip vortex persists downstream and strengthens with increasing Reynolds number. This is due to reduced viscous effects and the consequent increased circulation, which allows the vortex core to persist downstream. As was observed in the instantaneous flows, increased sweep weakens and breaks down the tip vortex, reflected by the shorter tip structures in the $\Lambda = 30^\circ$ and $45^\circ$ cases. We do, however, note that in the case $(Re,\Lambda) = (10000, 15^\circ)$, while the vortex appears to break down into smaller structures in the instantaneous flow field, it remains coherent in the time-averaged flow. 

For $Re\geq2500$, a region of elevated vorticity develops on the suction side near the trailing edge at all sweep angles, which can be seen from the Q-criterion isosurface $Qc^2/U_{\infty}^2=5$. In the unswept case, this strong vortical region is mainly located in the outboard, where the shear layer is deflected by the tip vortex. As the sweep angle increases, the high-vorticity region on the suction side of the wing moves toward the root.  This occurs where the shear layer developing above the wing rolls up due to the sweep induced spanwise velocity towards the tip. For $(Re,\Lambda) = (2500, 45^\circ)$, we observe the emergence of the aforementioned inboard vortical structure that persists into the wake far downstream. The increase in vorticity near the root, associated with the development of this elongated inboard vortical structure, occurs simultaneously with an extension of the reattachment region. 

For $\Lambda=45^\circ$, when the Reynolds number is increased from $Re=2500$ to $5000$, the vortical structures in the wake appear smaller and less persistent. In the time-averaged flow for $(Re,\Lambda) = (5000, 45^\circ)$, there appear to be two converging, elongated vortical structures emerging from the leading edge that persist into the wake. At $Re=10000$, there is instead a single, stronger, coherent, more persistent structure characteristic of the ram's horn vortex, which we will discuss in greater detail in Section~\ref{sec:rams}. At this Reynolds number, the reattached region at the root expands considerably, and the large, elongated vortical structure is clearly associated with the leading-edge vortex.

\begin{figure} 
\centering
\includegraphics[width=\textwidth]{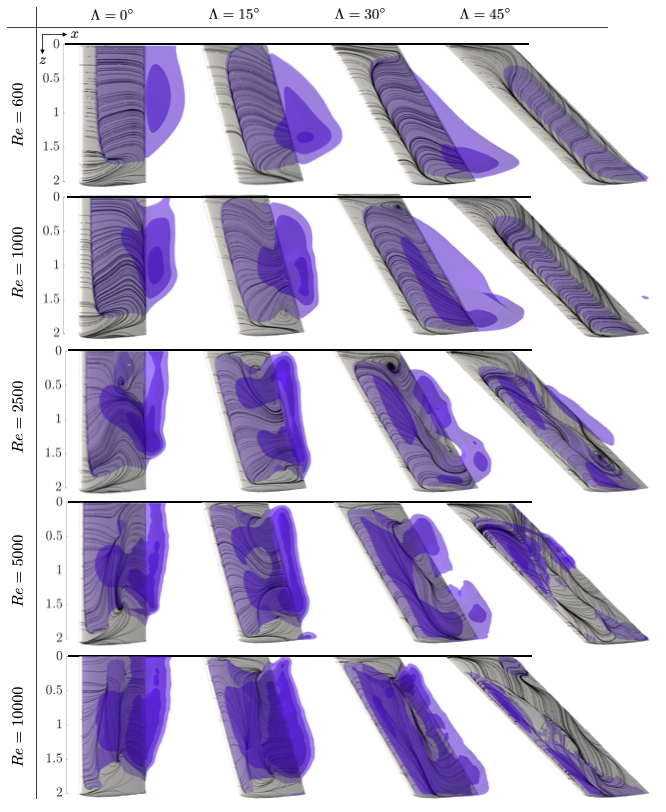}
\caption{Skin-friction lines and isocontours of streamwise average velocity $\bar{u}_x/U_\infty=0, -0.1$, and $-0.2$. \label{fig:RecirculatioTA}}
\end{figure}

We report the regions of flow separation in Figure~\ref{fig:RecirculatioTA}, where three isosurfaces of reversed streamwise velocity are visualized for each case. These isosurfaces highlight not only the extent of the separated region but also the relative intensity of the reversed flow, providing a qualitative indication of how strongly the recirculation develops as both $Re$ and $\Lambda$ vary. Additionally, we visualize the surface skin-friction lines to help characterize the behavior of the three-dimensional separation~\citep{hp1882crd, delery2001robert}.

Although there is some variability at the highest sweep angle cases, there is a general tendency for the separation region to move toward the tip as sweep increases. The onset of separation, marked by a spanwise-oriented separation line in the skin-friction curves~\citep{delery2001robert}, advances toward the leading edge as the Reynolds number increases at all sweep angles. Additionally, for all cases with low sweep ($\Lambda=0^\circ$ and $15^\circ$), we observe attached flow at the tip as a result of the tip vortex. On the other hand, an increased Reynolds number generally corresponds with a stronger recirculation region and increased three dimensionality of the flow.

At the lowest Reynolds number $Re = 600$, the recirculation region appears smooth and extends over most of the span. Towards the tip, the skin-friction lines from the leading edge start to turn towards the root, highlighting the influence of the tip vortex. When increasing sweep at this Reynolds number, the reversed flow region gradually shifts outboard towards the tip and there is a tip-ward turning of the skin friction lines near the root, indicative of the increasing spanwise advection induced by sweep, which is consistent with observations at higher Reynolds numbers by~\cite{neal2026three}. 
A similar trend is observed when increasing sweep for the $Re=1000$ cases, however, we start to see signs of three-dimensional deformation of the separated flow region.

At $Re = 2500$ and $5000$, as the reversed flow intensifies, more pronounced changes emerge in the topology of the recirculation region.
Notably, for $\Lambda = 15^\circ$ and $30^\circ$ at these Reynolds numbers, the high-intensity regions of the reversed flow start to divide into two spatially distinct inboard and outboard regions. This division arises from a change in the sign of the spanwise velocity along the wing, reflecting a balance between inboard-directed flow driven by tip effects and the outboard spanwise flow induced by sweep.

At higher Reynolds numbers ($Re = 5000$ and $10000$), the influence of sweep becomes more pronounced. For low sweep angles ($\Lambda \le 30^\circ$), the separated region still extends over much of the wing surface, but the reverse flow is more pronounced than in the lower Reynolds number cases. Additionally, a spanwise-oriented secondary recirculation region emerges near mid-chord, similar to observations in flows over periodic wings \citep{rolandi2025biglobal}. On the contrary, at the highest sweep angle $\Lambda = 45^\circ$, the flow exhibits substantially different separation behavior. At this sweep, a compact laminar separation bubble is formed close to the leading edge and wingtip, which is closely associated with the ram’s horn vortex.

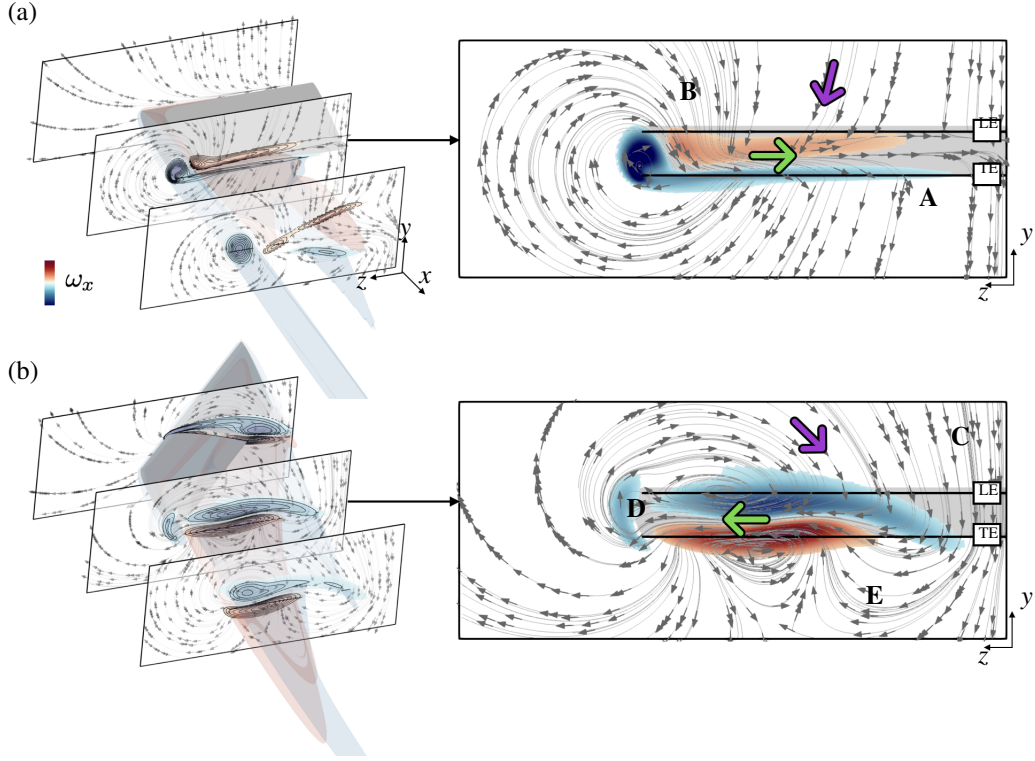
\begin{figure} 
\centering
\includegraphics[width=0.805\textwidth]{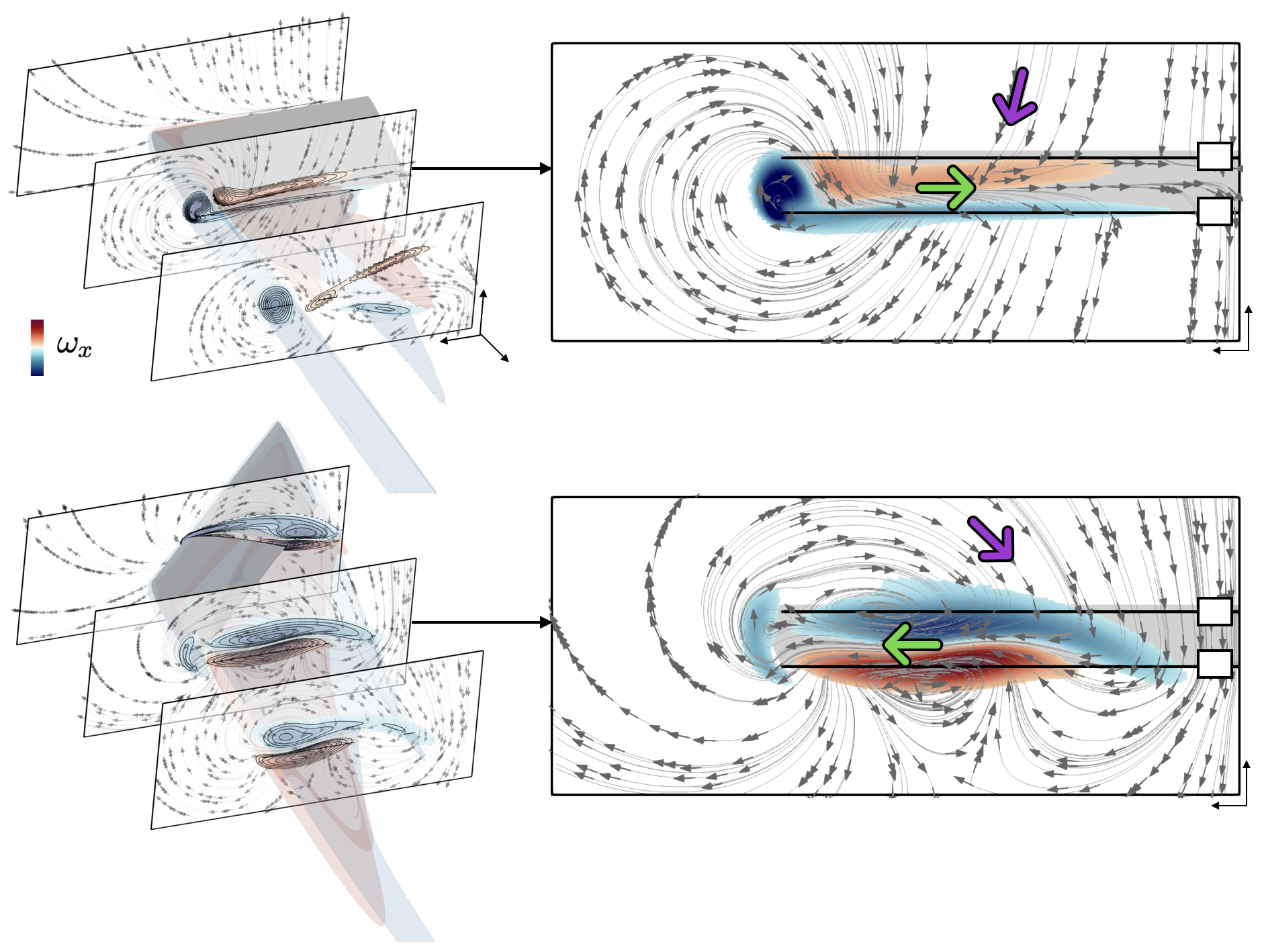}
\put(-383,280){(a)}
\put(-383,145){(b)}
\put(-236,199){{$y$}}
\put(-252,179){{$z$}}
\put(-228,181){{$x$}}
\put(-17.5,239.5){\tiny{LE}}
\put(-17.5,222.5){\tiny{TE}}
\put(-17.5,101){\tiny{LE}}
\put(-17.5,85){\tiny{TE}}
\put(-40,210){\textbf{A}}
\put(-130,250){\textbf{B}}
\put(-28,120){\textbf{C}}
\put(-60,60){\textbf{E}}
\put(-150,93){\textbf{D}}
\put(-2,197){{$y$}}
\put(-18,175){{$z$}}
\put(-2,60){{$y$}}
\put(-18,38){{$z$}}
\caption{Isosurfaces of time-averaged streamwise vorticity for $Re=600$ at (a) $\Lambda=0^\circ$ and (b) $\Lambda=45^\circ$. Contours of time-averaged streamwise vorticity and streamlines of $\bar{u}_{yz}=(0,\bar{u}_y,\bar{u}_z)$ are shown at $x=x_{\text{tip}}-0.5, x_{\text{tip}}+1$,and $x_{\text{tip}}+2.5$. Details of slices at $x=x_{\text{tip}}+1$ are shown on the right. Thick horizontal black lines indicate the projection of the leading edge (LE) and trailing edge (TE). \label{fig:VortXStreamlines}}
\end{figure}

To better understand the formation of the vortical structures as sweep is introduced, let us first examine the lowest Reynolds number case, $Re = 600$, to elucidate their development.
We show the time-averaged streamwise vorticity at $Re=600$ for $\Lambda=0^\circ$ and $45^\circ$ in Figure~\ref{fig:VortXStreamlines}. Streamlines of the flow projected onto slices parallel to the $yz$-plane are also shown.

For the unswept case (Figure~\ref{fig:VortXStreamlines}(a)), the slice at the trailing edge ($x=x_{\text{tip}}+1$) depicts an intensely concentrated region of negative streamwise vorticity corresponding to the tip vortex. The core is connected to a larger region of negative streamwise vorticity that forms along the spanwise direction from the pressure side, which lies below another elongated region of positive streamwise vorticity from the suction side. Moving to the downstream slice at $x=x_{\text{tip}}+2.5$, these two counter-rotating regions separate from each other vertically. In the $\Lambda=45^\circ$ case (Figure~\ref{fig:VortXStreamlines}(b)), the vorticity near the wing tip is noticeably lower than in the unswept configuration, reflecting a weakening of the tip vortex. The signs of the stratified layers of vorticity are reversed in comparison to the unswept case, with negative vorticity appearing over the wing due to the roll-up of the shear layer starting near the root.

Looking at the $x = x_{\text{tip}} + 1$ slice in greater detail (on the right in Figure~\ref{fig:VortXStreamlines}), we see that at $\Lambda=0^\circ$, the streamlines near the root (A) appear mostly vertical and the flow remains quasi two-dimensional. Moving towards the tip the flow is slightly deflected (purple arrow) due to the influence of the tip vortex. Some of the downwash from the tip vortex above the wing is pushed towards the root (green arrow) which visually demarcates the two opposite-signed vortical regions in the spanwise direction.

At $\Lambda = 45^\circ$, a significant region of negative spanwise velocity (purple arrow) develops above the wing as flow from the leading-edge rolls up in the yz-plane. This rolled up flow moves downward near the root (C), where part of it then moves outward toward the tip (green arrow). This outward spanwise motion opposes the flow around the wing tip, weakening the tip vortex (D). Below the wing, there is the interaction of several flow features: the tip flow, the counter-clockwise rotation corresponding to the pressure-side vorticity, and the downward flow from the root interacting with the clockwise rotation generated by the leading-edge shear layer roll-up. Notably, some of the flow going down from the root is deflected back upwards (E). It is this recirculating flow from the leading edge shear layer that eventually forms the Ram's horn vortex visible at higher Reynolds numbers.

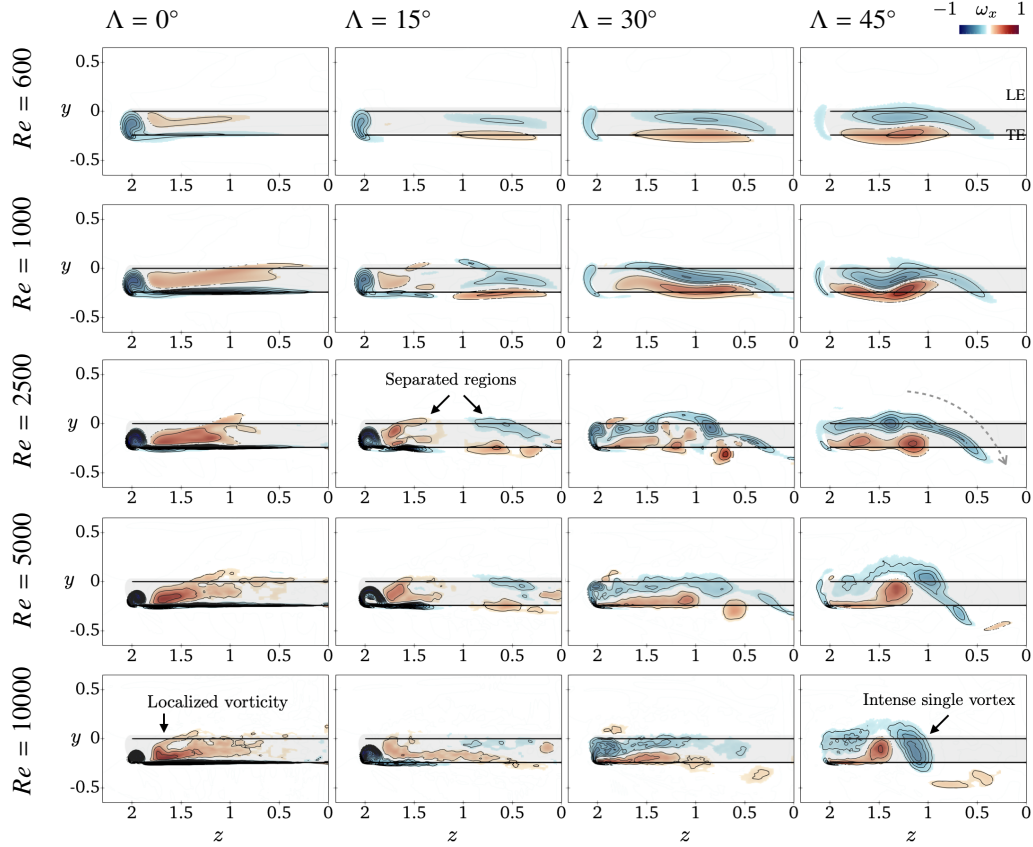
\begin{figure} 
\centering
\includegraphics[width=0.81\textwidth]{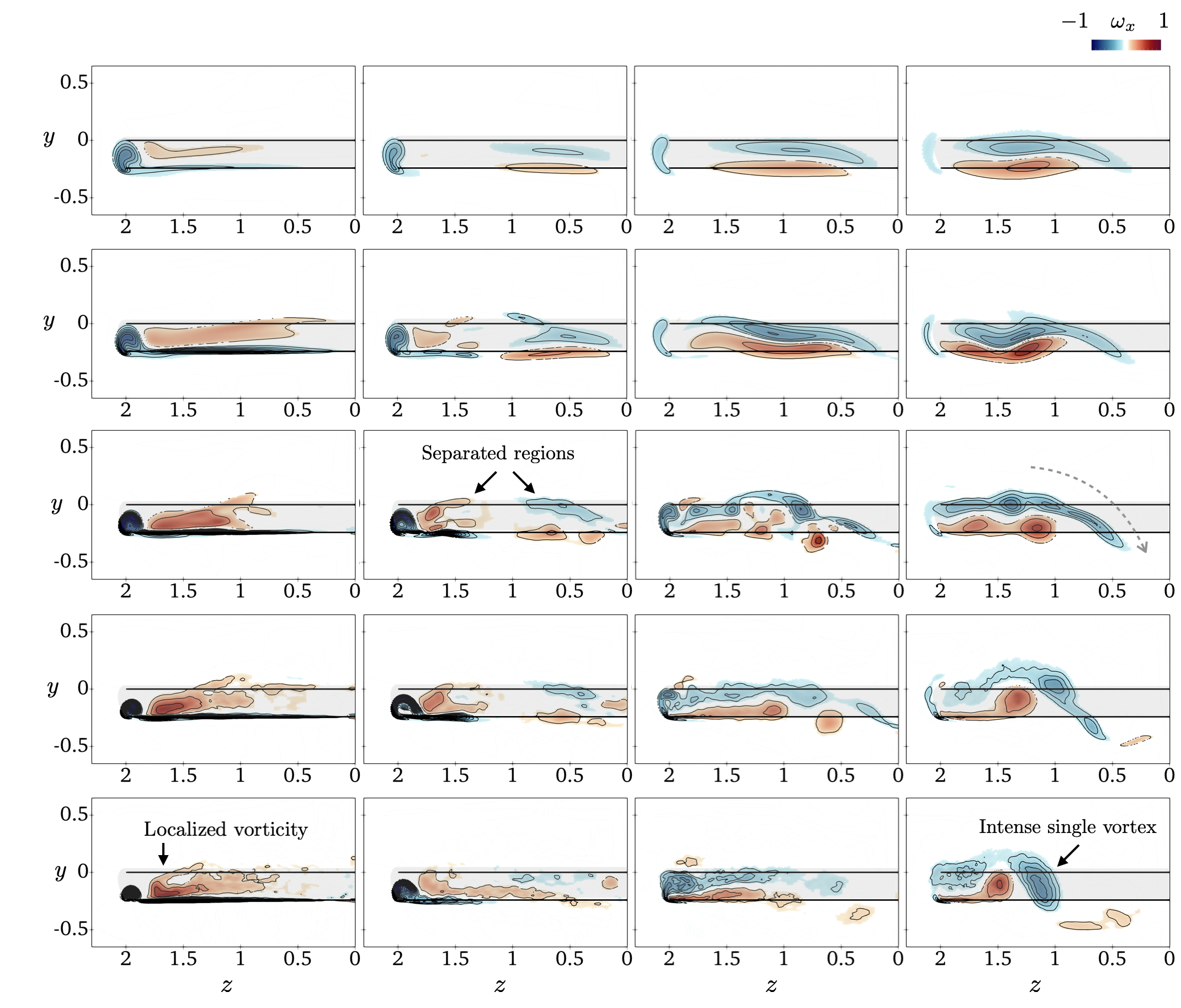}
\put(-385,265){\rotatebox{90}{$Re=600$}}
\put(-385,204){\rotatebox{90}{$Re=1000$}}
\put(-385,145){\rotatebox{90}{$Re=2500$}}
\put(-385,85){\rotatebox{90}{$Re=5000$}}
\put(-385,22){\rotatebox{90}{$Re=10000$}}
\put(-350,311){{$\Lambda=0^\circ$}}
\put(-260,311){{$\Lambda=15^\circ$}}
\put(-175,311){{$\Lambda=30^\circ$}}
\put(-85,311){{$\Lambda=45^\circ$}}
\put(-11,284){\tiny{LE}}
\put(-11,269){\tiny{TE}}
\caption{Contours of time-averaged streamwise vorticity at $x=x_{\text{tip}}+1$. Thick horizontal black lines indicate the projection of the leading and trailing edge, and the wing is shown in gray.  \label{fig:SliceXVort}}
\end{figure}

Now that we have described how the flow changes with sweep for the $Re=600$ case, we now present the contours of the time-averaged streamwise vorticity on the $yz$-plane at $x = x_\text{tip} + 1$ for the different Reynolds numbers and sweep angles in Figure~\ref{fig:SliceXVort}. 
In the unswept configuration, the tip vortex appears as negative streamwise vorticity aligned with the trailing-edge tip, strengthening with Reynolds number. Elongated layers of negative (trailing edge) and positive (leading edge) streamwise vorticity are also present across all Reynolds numbers, however their spatial extent becomes increasingly localized as Reynolds number increases. In the shear layer above the wing, increasing Reynolds number leads to a widening of the quasi two-dimensional region at the root (A in Figure~\ref{fig:VortXStreamlines}(a)) due to the increased momentum. The shear layer rolls up along the spanwise direction, thus the spanwise vorticity increases, while the streamwise vorticity decreases. This is reflected by the reduction of the spanwise extent of the region positive streamwise vorticity.

At $\Lambda = 15^\circ$, the streamwise-vorticity already exhibits significant changes from the unswept cases. The outward flow emanating from the root produces streamwise vorticity of opposite sign in the inboard region, as shown previously in Figure~\ref{fig:VortXStreamlines}. This results in two distinct spanwise regions of opposing vorticity, with an antisymmetric distribution in sign (e.g., the $(Re,\Lambda)=(1000,15^\circ)$ case). Further increases in sweep angle to $\Lambda = 30^\circ$ and $45^\circ$ coincide with the reversed-vorticity region expanding toward the wing tip. This trend reflects the strengthening sweep-induced spanwise flow, which increasingly dominates the near-wake vorticity distribution.

At $\Lambda = 45^\circ$ the strength of the tip vortex is considerably decreased in comparison with the unswept case. As Reynolds number increases at this sweep angle, the flow from the leading edge shear layer becomes increasingly rolled up by the spanwise flow. This causes the negative vorticity region above the wing to curl downward and become increasingly concentrated towards the outboard region of the wing, as indicated by the gray arrow at $Re=2500$ and $\Lambda=45^\circ$. At $Re = 5000$, the negative vorticity region begins to separate into two distinct vortical structures near the inboard and at $Re = 10000$, the inboard negative vorticity develops as a single, more intense vortex, which corresponds to the attached ram's horn vortex seen in Figure~\ref{fig:FlowField}.

\begin{figure} 
\centering
\includegraphics[width=0.815\textwidth]{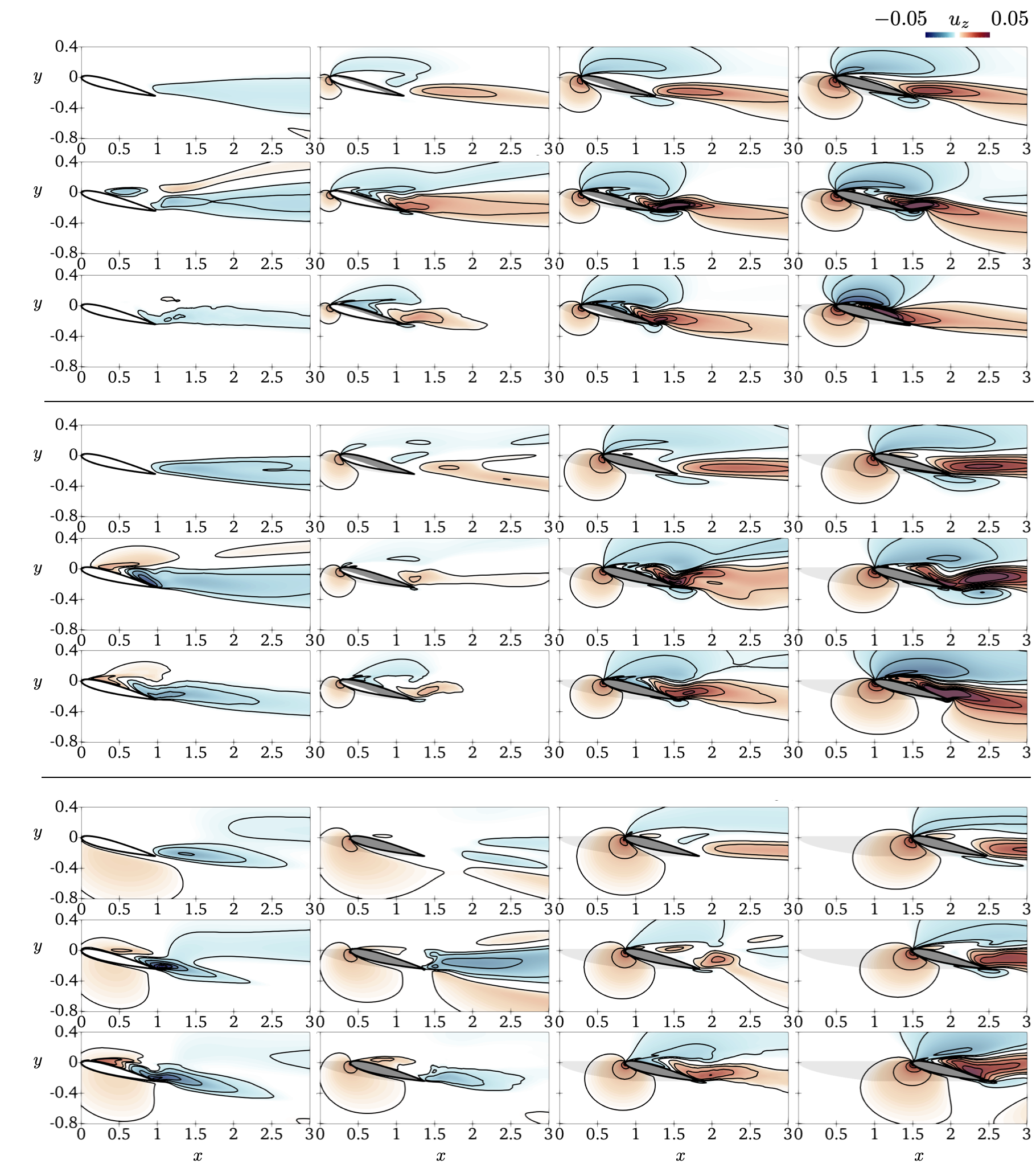}
\put(-380,420){\rotatebox{0}{(a)}}
\put(-380,275){\rotatebox{0}{(b)}}
\put(-380,135){\rotatebox{0}{(c)}}
\put(-353,385){\rotatebox{0}{$Re=600$}}
\put(-353,342){\rotatebox{0}{$Re=2500$}}
\put(-353,300){\rotatebox{0}{$Re=10000$}}
\put(-353,245){\rotatebox{0}{$Re=600$}}
\put(-353,202){\rotatebox{0}{$Re=2500$}}
\put(-353,162){\rotatebox{0}{$Re=10000$}}
\put(-353,103){\rotatebox{0}{$Re=600$}}
\put(-353,61){\rotatebox{0}{$Re=2500$}}
\put(-353,20){\rotatebox{0}{$Re=10000$}}
\put(-352,419){{$\Lambda=0^\circ$}}
\put(-262,419){{$\Lambda=15^\circ$}}
\put(-177,419){{$\Lambda=30^\circ$}}
\put(-88,419){{$\Lambda=45^\circ$}}
\caption{Contours of time-averaged spanwise velocity on $xy$-planes at (a) $z=0.5$, (b) $z=1.0$, and (c) $z=1.5$.  \label{fig:ZSlice_Uz}}
\end{figure}
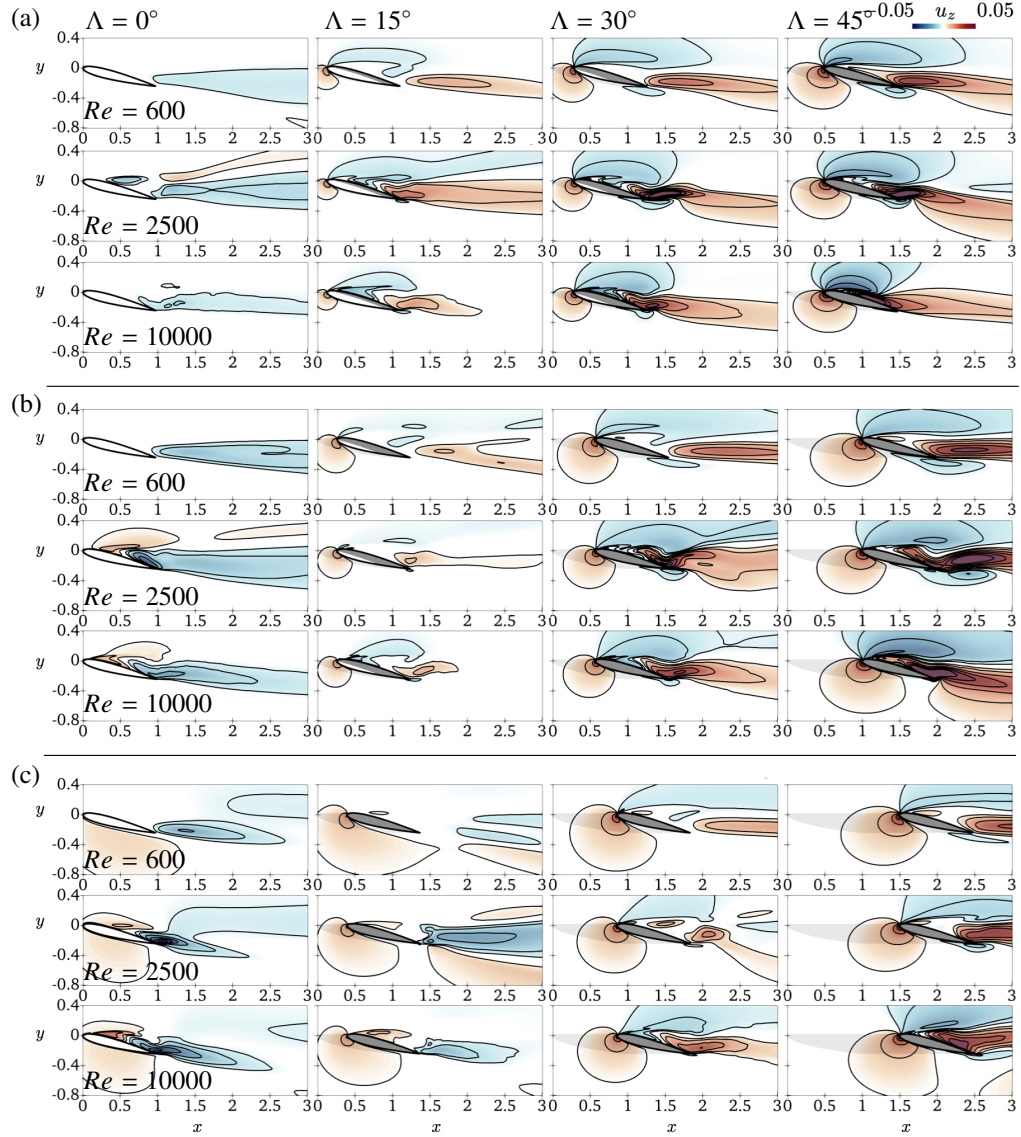

Although the change in vorticity sign with increasing sweep is most prominently observed at lower Reynolds numbers, the same mechanism also underpins the behavior at higher Reynolds numbers. Fundamentally, the insights gained from the low-Reynolds number cases allow us to intuit the trends in the vorticity distribution even at higher Reynolds numbers, where the effect is not as obvious. Overall, it is the change in spanwise velocity that drives the change in the vorticity distributions across all Reynolds numbers. In particular, there is a competition between the negative spanwise velocity that develops as a consequence of the downwash (green arrow in Figure~\ref{fig:VortXStreamlines}(a)) and the positive spanwise velocity that develops as a consequence of the wing sweep (green arrow in Figure~\ref{fig:VortXStreamlines}(b)).

To further illustrate this point, Figure~\ref{fig:ZSlice_Uz} presents contours of the time-averaged spanwise velocity, $\overline{u}_z$, on $xy$-planes at $z=0.5$, $1.0$, and $1.5$ for $Re=600$, $2500$, and $10000$. For the unswept wing ($\Lambda=0^\circ$), the wake is characterized by predominantly negative spanwise velocity, indicating flow directed toward the wing root. This behavior is observed at all spanwise stations and Reynolds numbers and results from the strong influence of the tip vortex, which entrains fluid inboard, consistent with the streamline patterns shown in Figure~\ref{fig:VortXStreamlines}. As the Reynolds number increases, localized regions of positive $\overline{u}_z$ appear near the upper shear layer, particularly at $z=1.0$ and $1.5$. These regions are associated with the increasing strength of the tip-vortex-induced downwash, which gets redirected toward the wing tip.

The introduction of sweep fundamentally modifies the spanwise velocity distribution. At $\Lambda=15^\circ$, positive $\overline{u}_z$ first appears near the root region, reflecting the onset of sweep-driven spanwise transport from root to tip. Consequently, the inboard flow observed in the unswept configuration is weakened and, in some regions, reversed. At the same time, the midspan section exhibits comparatively weak spanwise motion, while the outboard region remains influenced by the tip vortex and continues to exhibit substantial regions of negative $\overline{u}_z$. This indicates a competition between the sweep-induced outward transport and the tip-vortex-induced inward transport.

As the sweep angle increases to $\Lambda=30^\circ$ and $45^\circ$, the sweep-induced mechanism becomes increasingly dominant. Positive spanwise velocity occupies a progressively larger portion of the wake at all shown spanwise locations. The magnitude of $\overline{u}_z$ also increases with Reynolds number, demonstrating that the spanwise flow intensifies as inertial effects become more pronounced. Furthermore, the location of the maximum spanwise velocity shifts upstream with increasing Reynolds number, indicating that the sweep-induced transport of momentum occurs closer to the wing. Overall, these observations further reinforce that sweep progressively transforms the wake from a tip-vortex-dominated flow characterized by inward spanwise motion to a sweep-dominated flow exhibiting strong outward transport toward the wing tip.

\subsection{The ram's horn vortex}\label{sec:rams}

\begin{figure} 
\centering
\includegraphics[width=0.85\textwidth]{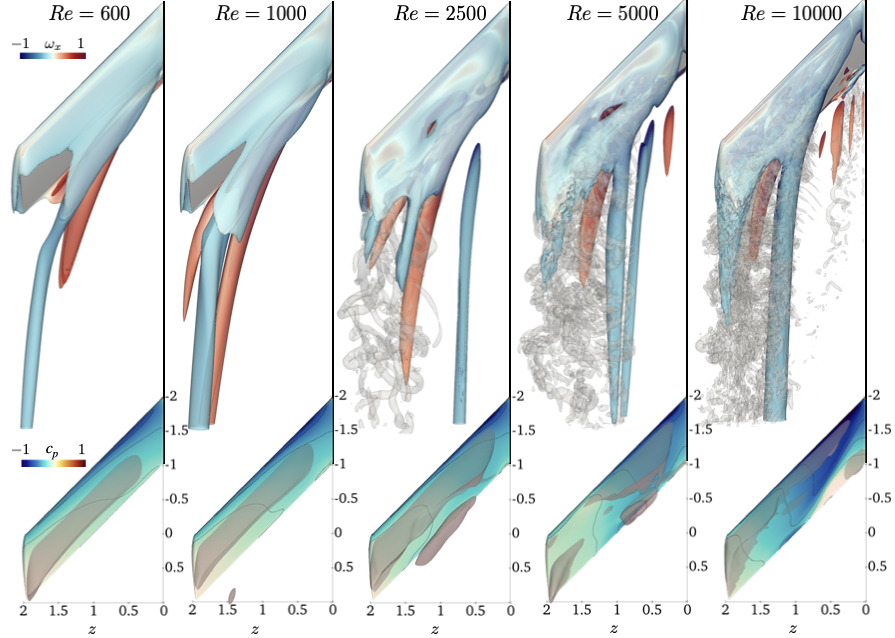}
\caption{Flow fields at $\Lambda=45^\circ$ across different Reynolds numbers. The top row depicts a visualization with the isosurface of time-averaged Q-criterion $Qc^2/U_{\infty}^2=0.5$ colored by time-averaged streamwise vorticity, superposed on translucent isosurface of the instantaneous Q-criterion. The bottom row depicts the pressure coefficient over the suction surface with isocontours shown as black lines. The recirculation region $\bar{u}_x=0$ is visualized as the pink translucent isosurface.\label{fig:Lambda45}}
\end{figure}

Ram's horn vortices have been reported over a wide range of Reynolds numbers. At high Reynolds numbers ($Re \geq 5\times10^5$), they are commonly associated with flow reattachment and root- or tip-stall phenomena \citep{black1956flow,poll1986spiral}, while more recent studies have shown that similar structures can also arise in low-Reynolds-number flows \citep{zhang2020laminar,burtsev2022linear}. Figure~\ref{fig:Lambda45} illustrates the evolution of the ram's horn vortex with Reynolds number. The vortical structures are visualized using isosurfaces of the $Q$-criterion colored by the streamwise vorticity.

As discussed in Figure~\ref{fig:VortXStreamlines}(b), regions of negative streamwise vorticity correspond to fluid originating from above the wing, primarily associated with the roll-up of the separated shear layer and the wingtip flow. In contrast, positive streamwise vorticity is associated with fluid originating from the pressure side of the wing, particularly from the trailing-edge vortex. The interaction of these two counter-rotating vortical structures gives rise to the characteristic ram's horn vortex.

Figure~\ref{fig:Lambda45} shows the evolution of the flow field at $\Lambda=45^\circ$ with increasing Reynolds number. Additionally, we show the pressure coefficient over the surface, given by
\[
c_p=\frac{\bar{p}-p_\infty}{0.5\rho_\infty U_\infty^2},
\]
where $\bar{p}$ is the time-averaged surface static pressure and $p_\infty$ is the freestream static pressure. Here, $0.5\rho_\infty U_\infty^2$ denotes the freestream dynamic pressure used for nondimensionalization. At $Re=600$, the ram's horn vortex forms from the roll-up of the separated shear layer approximately one chord length downstream of the leading edge. The vortex initially develops near the root, extends toward the tip, and is subsequently aligned with the streamwise direction. The recirculation region, identified by $\bar{u}_x=0$, is relatively uniform along the span, with attached flow at the root region.

This relatively homogeneous separation pattern is likewise reflected in the surface pressure distribution. At the low Reynolds numbers of $Re=600$ and $1000$, the pressure coefficient varies only weakly in the spanwise direction. The limited spanwise variation in $c_p$ further confirms the near-uniform nature of the separation pattern. Instead, the pressure distribution is primarily governed by the chordwise pressure gradient, with low-pressure values near the leading edge gradually recovering toward the trailing edge.

At $Re=1000$, the ram's horn vortex becomes stronger as both the rolled-up shear-layer vortex and the trailing-edge vortex intensify. Concurrently, the recirculation region shifts upstream toward the leading edge, indicating earlier boundary-layer separation. A similar Reynolds-number dependence was reported by \citet{rolandi2025biglobal} for a periodic swept-wing configuration.

For the $Re=2500$ case, the overall vorticity over the wing increases substantially, as evidenced by the time-averaged $Q$-criterion isosurface extending across the entire span. At the same time, the onset of unsteadiness alters the wake organization: the coherent vortical structures observed in the outboard wake at lower Reynolds numbers become less persistent and less pronounced. Instead, a new vortical structure emerges in the inboard region. This structure is characterized by negative streamwise vorticity, indicating its origin from the roll-up of the inboard shear layer. 

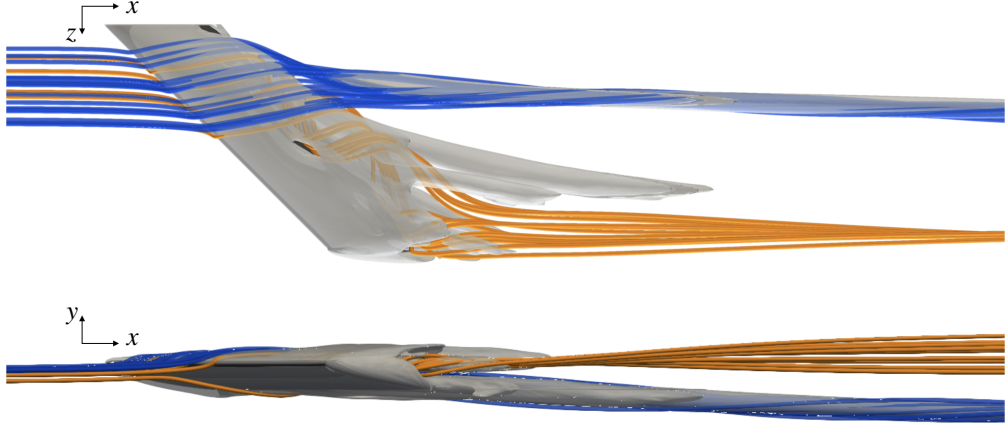
\begin{figure} 
\centering
\includegraphics[width=0.82\textwidth]{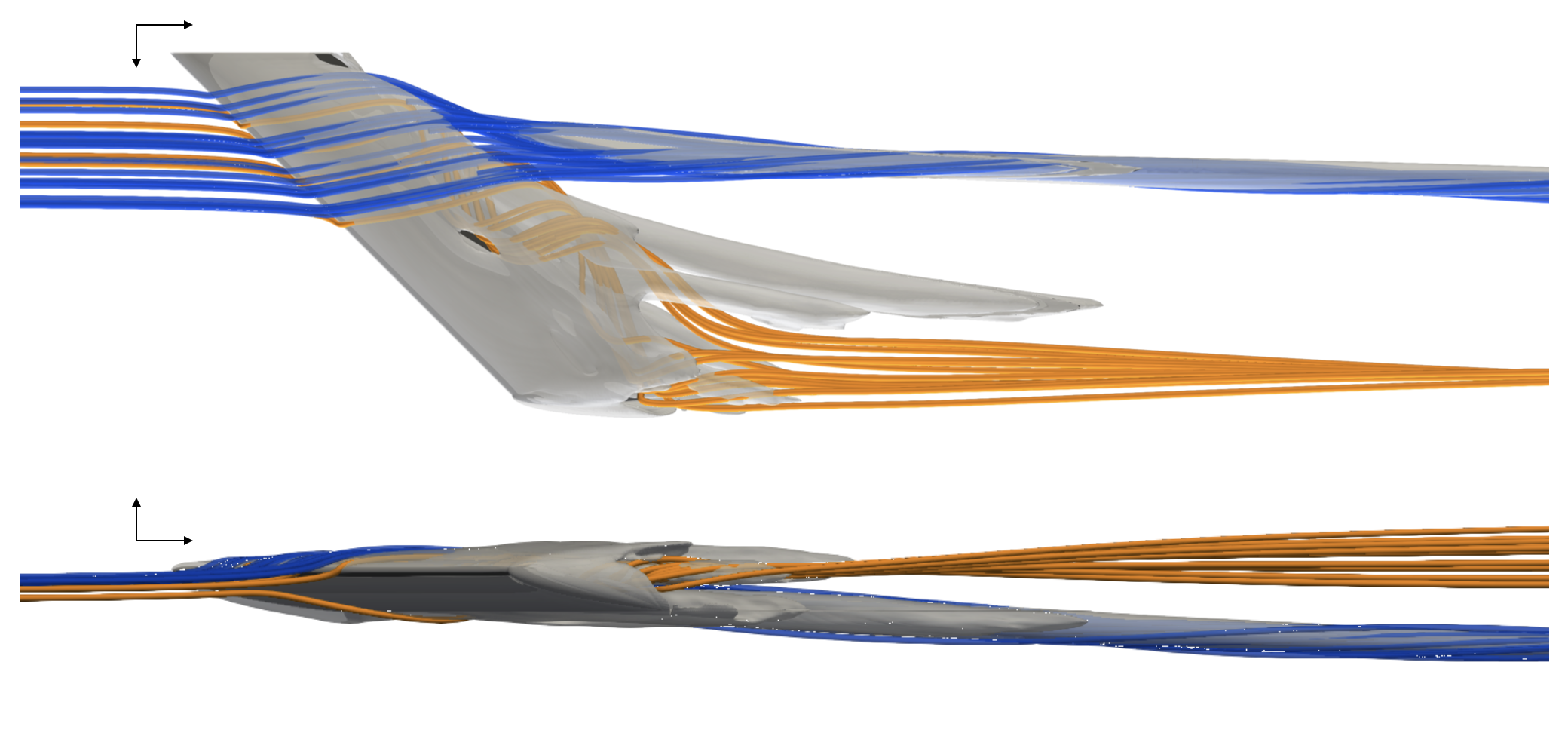}
\put(-358,165){{$z$}}
\put(-335,175){{$x$}}
\put(-358,60){{$y$}}
\put(-335,50){{$x$}}
\caption{Streamlines of time-averaged flow at $Re=2500$ and $\Lambda=45^\circ$ visualizing the inboard vortical structure (blue) and shear layer roll-up (orange) superposed on translucent isosurface of Q-criterion. \label{fig:InboardVortex}}
\end{figure}

The formation of the so-called inboard vortex can be traced to a portion of the root shear layer that is not immediately rolled up by the sweep-induced spanwise flow. Rather, it initially remains aligned with the streamwise direction. However, as it convects downstream, this flow interacts with positive spanwise velocity in the wake underneath it, which twists the streamwise-oriented flow, leading to the formation of this  inboard vortical structure. A detailed visualization of this mechanism is provided in Figure~\ref{fig:InboardVortex}.

These observations are consistent with high-Reynolds-number delta-wing flows, where a persistent inboard leading-edge vortex has also been reported \citep{furman2013turbulent,luckring2019discovery}. \citet{zhang2020laminar} further identified alternating regions of positive and negative streamwise vorticity, dubbed finger-like structures, in steady flow over finite swept wings at $Re=400$, observed for $sAR>2$ and $\Lambda=45^{\circ}$. While the spacing between neighboring streamwise vortices remains constant with increasing $sAR$ at fixed Reynolds number \citep{zhang2020laminar}, it varies with both angle of attack and Reynolds number in delta-wing flows \citep{furman2013turbulent}, consistent with our findings.

As the Reynolds number increases, the primary vortical structure that persists into the wake moves away from the tip towards the midspan. In our case, between $Re=2500$ and $5000$, the two elongated negative streamwise vortical structures originating over the wing move closer together. Over this range, the shear layer rolls up progressively nearer the leading edge at the root, coinciding with a strengthening and rightward shift of the negative streamwise vortex aligned with $z=1.5$ at $Re=2500$. This evolution ultimately leads to the formation of the attached inboard ram’s horn vortex seen in the $Re=10000$ case. Indeed, at $Re = 10000$, there is a single, well-defined vortex that forms from the roll-up of the root shear layer.

This behavior is accompanied by significant changes in the surface-flow topology. For the $Re=10000$ case in Figure~\ref{fig:Lambda45}, the separation line is shifted upstream toward the leading edge. Consistently, the pressure field shows an enlarged low-$c_p$ region near the root, along with increasingly pronounced spanwise variations in pressure coefficient. These changes stem from a growing contribution from vortex lift, as the rolled-up shear-layer vortex is reattached to the suction surface and strengthens local lift. In addition, a “conical” vortical structure forms near the tip, associated with separated flow in that region, consistent with the observations of tip stall by \citet{neal2026three}.

Overall, Figure~\ref{fig:Lambda45} illustrates the evolution of the ram’s horn vortex with Reynolds number. A distinct ram’s horn vortex forms at low $Re$. As $Re$ increases, inboard vortical structures and multiple trailing-edge vortices emerge, likely driven by mechanisms similar to those responsible for the formation of finger-like and inboard vortices \citep{jardin2017coriolis,zhang2020laminar}. At the highest Reynolds numbers, a well-defined ram’s horn vortex reappears, accompanied by a persistent tip vortex.

\begin{figure} 
\centering
\includegraphics[width=\textwidth]{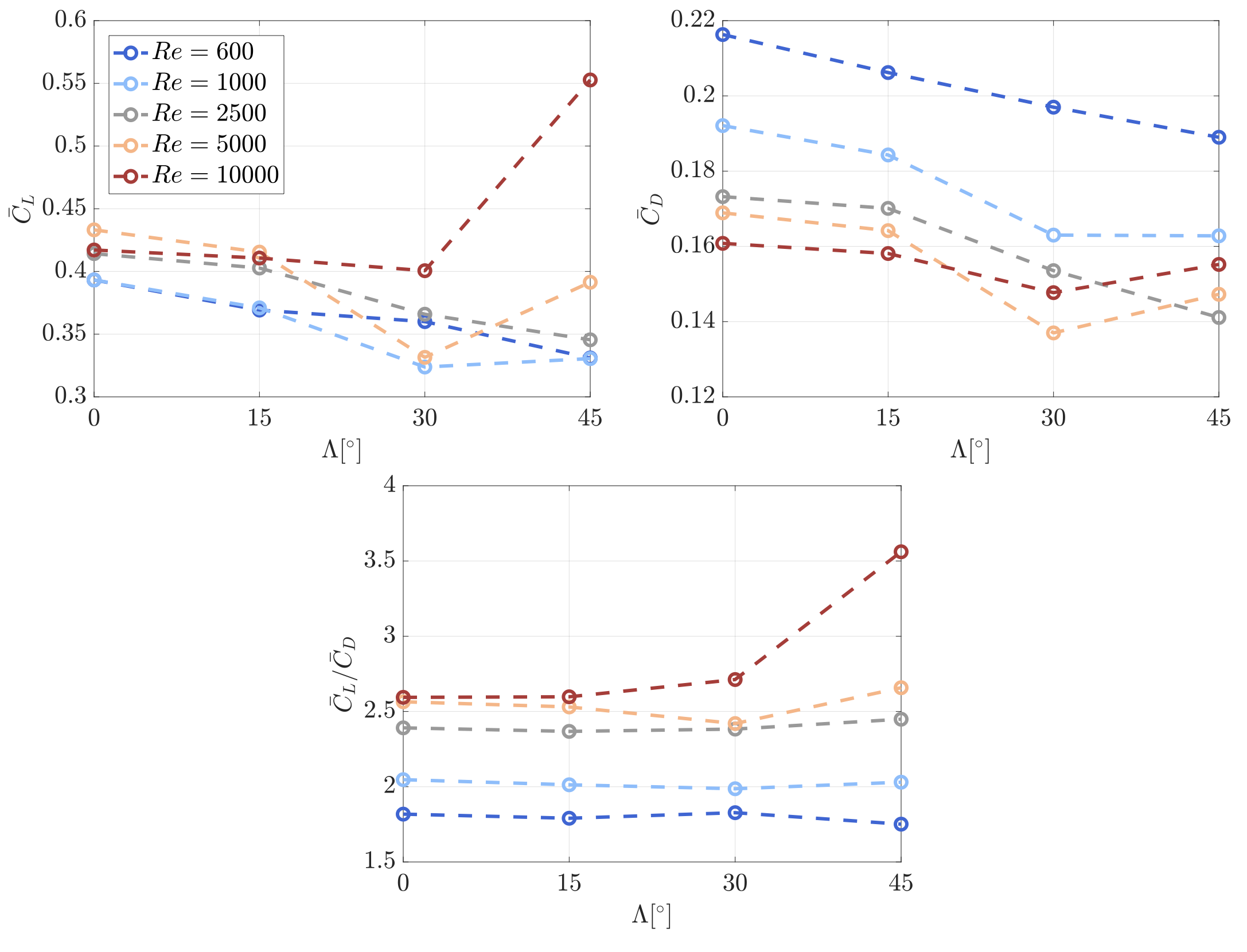}
\caption{Mean lift coefficient ($\bar{C}_L$), mean drag coefficient ($\bar{C}_D$) and lift-to-drag ratio ($\bar{C}_L/\bar{C}_D$). \label{fig:CDCL}}
\end{figure}

\subsection{Aerodynamic coefficients}\label{sec:AeroCoeff}
In Section \ref{sec:InstantaneousFlow}, we analyzed the time history of the aerodynamic coefficients, focusing on the amplitude and period of their oscillations (Figure~\ref{fig:Forces}). We now turn to their mean values, reported in Figure~\ref{fig:CDCL}, as a function of sweep angle for the different Reynolds numbers.

The mean lift coefficient generally decreases almost monotonically with sweep at low Reynolds numbers. However, there is a noticeable change in behavior as Reynolds number increases. At $Re=5000$, $\bar{C}_L$ initially decreases with sweep before recovering at $\Lambda=45^\circ$. However, at $Re=10000$ $\bar{C}_L$ remains fairly constant before rising sharply at $\Lambda=45^\circ$, even exceeding the lift of the baseline unswept case. The drag coefficient follows a similar trend. The value of $\bar{C}_D$ decreases monotonically for $Re \leq 2500$, but increases again at the highest sweep angles for $Re=5000$ and $10000$. The lift-to-drag ratio ($\bar{C}_L/\bar{C}_D$) remains mostly constant with respect to sweep angle for $Re\leq2500$. At higher-Reynolds numbers, we see increased aerodynamic efficiency at $\Lambda=45^\circ$ due to vortex lift.

\begin{figure} 
\centering
\includegraphics[width=0.83\textwidth]{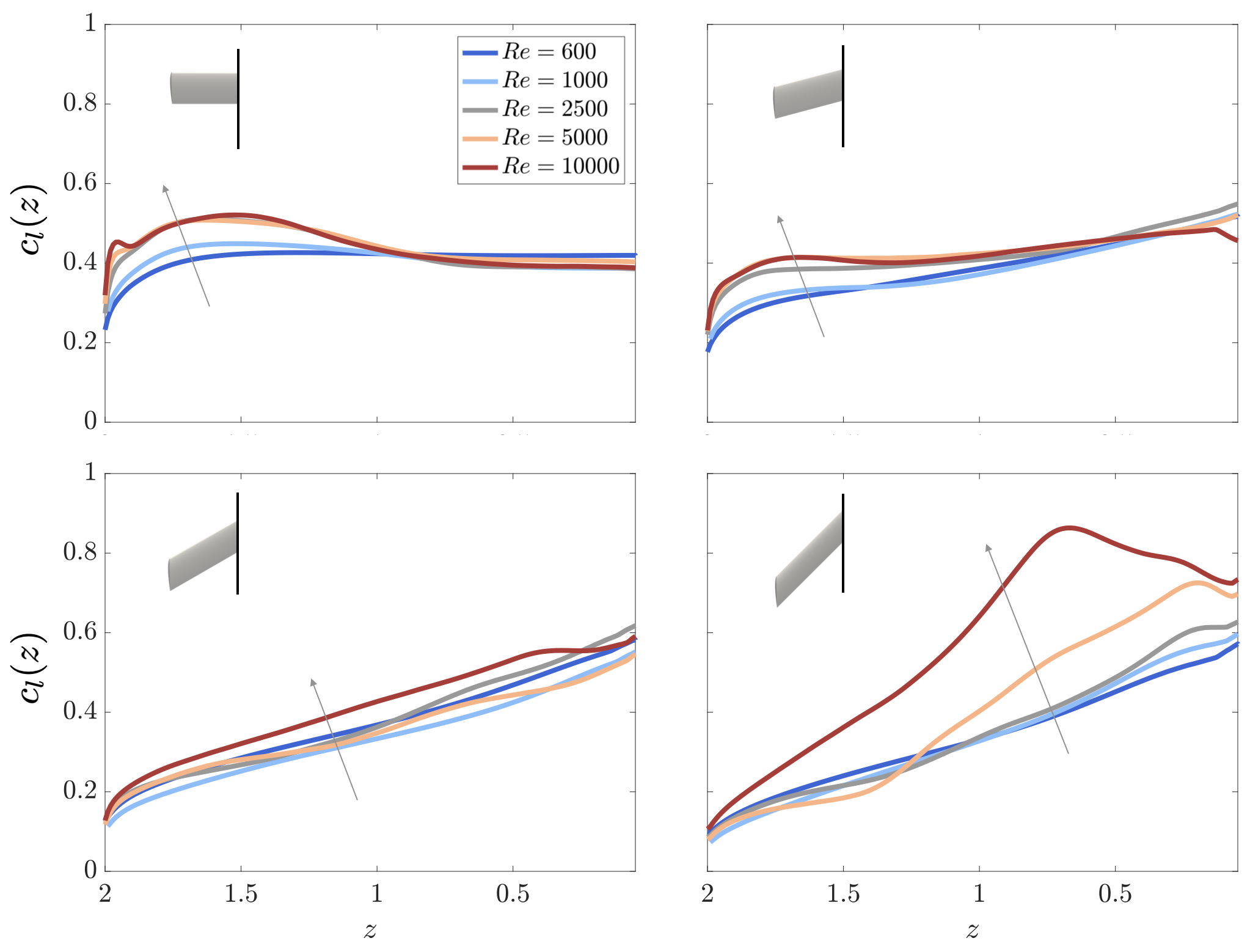}
\put(-350,277){$\Lambda=0^\circ$}
\put(-163,277){$\Lambda=15^\circ$}
\put(-350,138){$\Lambda=30^\circ$}
\put(-163,138){$\Lambda=45^\circ$}
\put(-323,185){Increasing $Re$}
\caption{Sectional lift coefficient varying the Reynolds number for each sweep angle considered  \label{fig:SectLift}}
\end{figure}
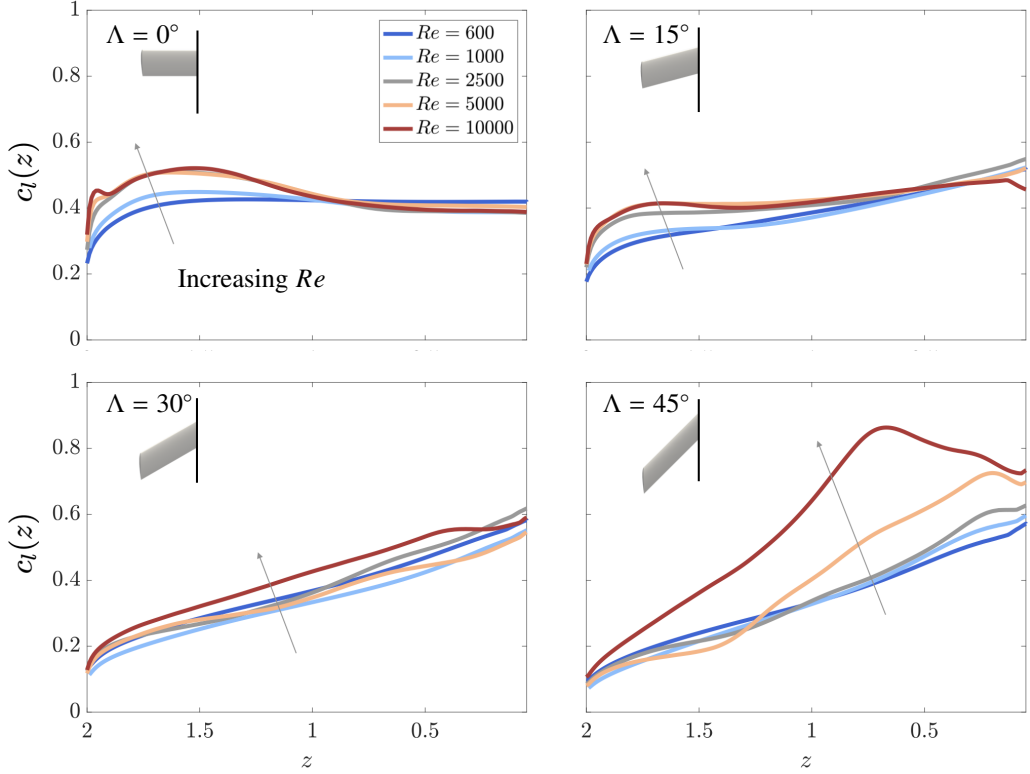

To further understand the mechanisms behind the increase in mean lift, we examine the spanwise distribution of the sectional lift coefficient, as shown in Figure~\ref{fig:SectLift}. The sectional lift coefficient $\bar{c}_l(z)$ is defined based on the pressure difference integrated over the airfoil surface at each spanwise location $z$:
\begin{equation}
\bar{c}_l(z) =
\frac{1}{\tfrac{1}{2}\rho_\infty U_\infty^2}
\int_{S(z)} \left[\bar{p}(x,y,z) - p_\infty \right] \, \mathrm{d}S,
\end{equation}
where $\mathrm{d}S$ denotes the differential surface measure on the airfoil section $S(z)$ at a given spanwise location $z$ and $\bar{p}(x,y,z)$ is the time-averaged static pressure on the surface.

At $\Lambda=0^\circ$ the sectional lift is nearly uniform along the span for $Re=600$ and $1000$, decreasing at the tip, consistent with \cite{smith2024effect}. For $Re \geq 2500$, however, the outer half of the wing shows a clear increase in lift, likely linked to the localized vorticity discussed in Figure~\ref{fig:SliceXVort}. Although a uniform lift distribution typically minimizes induced drag \citep{gudmundsson2013general}, here the drag remains higher at low Reynolds numbers due to viscous effects (see Figure~\ref{fig:CDCL} at $\Lambda=0^\circ$).

As sweep is introduced, the sectional lift increases near the root while decreasing toward the tip. In general, increasing the Reynolds number leads to an increase of sectional lift along the span. Notable is the $(Re,\Lambda) = (10000, 45^\circ)$ case, which exhibits a substantial lift enhancement near the root. This behavior is associated with the formation of a ram’s horn vortex, accompanied by reattachment of the rolled-up leading-edge flow in the root region. The high velocity within this leading-edge vortex induces a low-pressure region over the upper surface of the wing, thereby increasing the sectional lift coefficient \citep{vos2015aerodynamics}.

\section{Conclusions}\label{sec:Conclusions}
In this work, we have examined the wake dynamics and characteristics of the flow around a finite wing of $sAR=2$ at a $14^\circ$ angle of attack. The analysis was performed at different wing sweep angles, $\Lambda=0^\circ,15^\circ,30^\circ$ and $45^\circ$, and over transitional Reynolds numbers. In particular, we focused on the changes of the flow characteristics from low Reynolds numbers $Re\approx 10^2$ to high Reynolds numbers $Re\approx 10^4$. Our findings elucidate the transition from low Reynolds numbers, where the sweep decreases the lift coefficient while contributing to separation at the tip, and moderate Reynolds numbers, where the leading edge vortex reattaches at the root, generating the so-called vortex lift.  

In particular, we clarified the mechanisms responsible for the changes in vorticity distribution that arise with increasing sweep angle, and demonstrated how these modifications in the flow field lead to the formation of the leading-edge vortex. A key observation is the development and intensification of vortical structures in the inboard region. This structure is first observed with pressure variations at $Re=600$ and $1000$, and clearly visible as a vortical structure at $Re=2500$. This vortical structure is the low Reynolds number representation of the leading edge vortex observed at $Re=10000$.

Interestingly, the $\Lambda=15^\circ$ and $\Lambda=30^\circ$ cases exhibit intermediate flow characteristics: the outboard region behaves similarly to the unswept configuration, while the inboard region displays features more consistent with the highly swept case. This contrast is driven by the spanwise velocity distribution along the wing, which changes sign across the span when transitioning from the unswept case (dominated by negative spanwise velocity) to the swept case (dominated by positive spanwise velocity).

We also characterized the ram's horn vortical structure, which emerges at the highly swept case of $\Lambda=45^\circ$, over the Reynolds number. For the considered aspect ratio, at the lowest Reynolds numbers, the ram's horn vortex emerges as a distinct structure that forms from the shear layer roll-up above the wing, extends from the root toward the tip and aligns gradually with the streamwise direction. As $Re$ increases to $2500$ and $5000$, the flow exhibits an increased level of three dimensionality. A vortical structure emerges in the inboard region, due to the tilting of the outer part of the shear layer above the wing as it develops downstream. This stage marks the transition from a simple, tip-aligned ram’s horn vortex to a more complex vortical flow. The emergence of the inboard vortical structure was interpreted as arising from a mechanism analogous to the formation of the finger-like vortices observed in low-Reynolds-number flows \citep{zhang2020formation}. Finally, by $Re=10000$, the time-averaged flow reveals a single, well-defined leading-edge vortex near the root and a persistent tip vortex. The ram’s horn vortex thus evolves from a tip-oriented structure at low $Re$ to a reattached, inboard vortex at higher $Re$. These observations suggest that as $Re$ increases, the ram’s horn and tip vortices progressively interact, indicating their eventual merging at even higher Reynolds numbers, consistent with the tip-stall associated with the ram’s horn vortex described by \citet{black1956flow} at $Re=500000$.

This work provided a systematic analysis of swept wings at transitional Reynolds numbers, contributing
to a deeper understanding of swept-wing wake dynamics, which is crucial for aerospace vehicle
design.

\section*{Acknowledgements}
This work was supported in part by high-performance computing time and resources from the DoD High Performance Computing. We also thank Luke Smith for his assistance in generating many of the meshes used in this study.
\section*{Funding}
This work was supported by the U.S. Air Force Office of Scientific Research (Grant Number FA9550-21-1-0174) and the U.S. Army Research Office (Grant Number W911NF-21-1-0060).

\bibliographystyle{jfm}
\bibliography{jfm}

\end{document}